\documentclass[conference,letterpaper]{IEEEtran}

\usepackage[utf8]{inputenc} 
\usepackage[T1]{fontenc}
\usepackage{fancyhdr}
\usepackage{lastpage}
\usepackage{url}
\usepackage{ifthen}
\usepackage{cite}
\usepackage[cmex10]{amsmath}
\usepackage{amsmath}
\usepackage{amssymb, amsfonts, mathrsfs}
\usepackage{amsthm}
\usepackage{bbm, dsfont}
\usepackage{bbold}
\usepackage{graphicx, url, color}
\usepackage{bbm}
\newtheorem{theorem}{Theorem}
\newtheorem{lemma}{Lemma}

\newcommand{\remove}[1]{}

\begin{document}
\title{Polar Coding for Multi-level 3-Receiver Broadcast Channels} 

% %%% Single author, or several authors with same affiliation:
% \author{%
%   \IEEEauthorblockN{Stefan M.~Moser}
%   \IEEEauthorblockA{ETH Zürich\\
%                     ISI (D-ITET)\\
%                     CH-8092 Zürich, Switzerland\\
%                     Email: moser@isi.ee.ethz.ch}
% }

%%% Several authors with up to three affiliations:
\author{%
  \IEEEauthorblockN{Karthik Nagarjuna Tunuguntla}
  \IEEEauthorblockA{
                    CMRR, University of California, San Diego\\
                    Email: tkarthik@eng.ucsd.edu}
  \and
  \IEEEauthorblockN{Paul H. Siegel}
  \IEEEauthorblockA{CMRR, University of California, San Diego\\
                    Email: psiegel@ucsd.edu}
}

%%% Many authors with many affiliations:
% \author{%
%   \IEEEauthorblockN{Albus Dumbledore\IEEEauthorrefmark{1},
%                     Olympe Maxime\IEEEauthorrefmark{2},
%                     Stefan M.~Moser\IEEEauthorrefmark{3}\IEEEauthorrefmark{4},
%                     and Harry Potter\IEEEauthorrefmark{1}}
%   \IEEEauthorblockA{\IEEEauthorrefmark{1}%
%                     Hogwarts School of Witchcraft and Wizardry,
%                     1714 Hogsmeade, Scotland,
%                     \{dumbledore, potter\}@hogwarts.edu}
%   \IEEEauthorblockA{\IEEEauthorrefmark{2}%
%                     Beauxbatons Academy of Magic,
%                     1290 Pyrénées, France,
%                     maxime@beauxbatons.edu}
%   \IEEEauthorblockA{\IEEEauthorrefmark{3}%
%                     ETH Zürich, ISI (D-ITET), ETH Zentrum, 
%                     CH-8092 Zürich, Switzerland,
%                     moser@isi.ee.ethz.ch}
%   \IEEEauthorblockA{\IEEEauthorrefmark{4}%
%                     National Chiao Tung University (NCTU), 
%                     Hsinchu, Taiwan,
%                     moser@isi.ee.ethz.ch}
% }

\maketitle

%%%%%%
%% Abstract: 
%% If your paper is eligible for the student paper award, please add
%% the comment "THIS PAPER IS ELIGIBLE FOR THE STUDENT PAPER
%% AWARD." as a first line in the abstract. 
%% For the final version of the accepted paper, please do not forget
%% to remove this comment!
%%
\begin{abstract}
   We consider achieving the rates in the capacity region of a multi-level 3-receiver broadcast channel, in which the second receiver is degraded with respect to the first receiver, with degraded message sets. The problem is to transmit a public message intended for all three receivers and a private message intended for the first receiver. Our interest in coding for the broadcast channel problem is due to a file transfer application in a client-server network  which has three clients, where this problem of broadcast channel can be applied.  We propose a two-level chaining strategy based on polar codes that achieves the capacity region of the considered setting without time-sharing. We also look at a slight variation of this problem, where the first receiver only requires to decode its own private message and the other two receivers require to decode another private message common to them. We observe that the capacity
   region does not enlarge and so the proposed polar coding strategy achieves the capacity region for this problem as well.

   %We do a rate-splitting of the private message intended for the first receiver to  implement the coding scheme that achieves the rates in the capacity region.
\end{abstract}

%% The paper must be self-contained. However, if you are referring to
%% a full version for checking certain proofs, please provide the
%% publically accessible location below.  If the paper is completely
%% self-contained, you can remove the following line from your
%% submission.
\begin{section}{Introduction}
\subsection{Background}
Arikan~\cite{arikan} constructed   capacity-achieving polar codes for binary input symmetric channels. Since then, many coding strategies have been introduced for multi-user settings using the polarization method~\cite{arikan2}, \cite{korada}, \cite{arikan3}.  Goela,  Abbe and Gastpar~\cite{goela}
introduced polar codes for $m$-user deterministic broadcast channels. They also  introduced polar coding for 2-user noisy broadcast channels. They implemented superposition and Marton schemes which involve some assumptions of degradation on the channel parameters to align the polar indices.  Mondelli, Hassani,  Sason, and Urbanke~\cite{mondelli} proposed schemes to remove such constraints using a polar-based chaining construction~\cite{hassani}, \cite{hassani2}. Chou and Bloch~\cite{chou} proposed a polar coding scheme for a broadcast channel with confidential messages. Alos and Fonollosa~\cite{olmo} proposed a polar coding scheme for a broadcast channel with two legitimate receivers, that receive a confidential and private message, and one eavesdropper.

In this paper, we consider the problem of achieving the rates in the capacity region for a memoryless(DM) $3$-receiver broadcast channel with degraded message sets~\cite{nair}~\cite{elgamal}. The second receiver is degraded with respect to the first receiver. The problem is to transmit a public message intended for all three receivers and a private message intended for the first receiver. Our motivation to consider this problem for the broadcast channel is due to a very useful practical file transfer application in a client-server network. We describe the file transfer application in a  client-server network, where our problem setting is applied,  in the following sub-section.

\subsection{Motivation with a client-server model} 

We consider a client server model, in which server sends data to its three clients that are computer, phone-$1$ and phone-$2$. Computer and phone-1 receive data from the server directly via internet. Phone-2 receives data from server indirectly through bluetooth connection between computer and phone-2. Computer supports both video and audio applications whereas the two phones only support audio application. Suppose that server has one audio and one video file to send its clients.  Note that all three clients here are interested in receiving the audio file. Also notice that computer is the only client interested in receiving both the audio and video files. This is shown in the Fig.\ref{fig3}.  In some scenarios, computer may just want to receive the video file.

The internet link and bluetooth link add noise to the signal received at the clients. So this can be modelled as noisy broadcast channel with 3-receivers which we are interested to look at in this paper. The goal is to find a method where the server can send both the audio and video files to its clients reliably at all possible data rates that can be supported by the network. It amounts to finding a coding scheme that can achieve rates in the capacity region of this broadcast channel problem.   This particular client-server setting with the file transfer scenario is highly applicable to house-hold internet links, which is why we are motivated to consider this problem. 

We define the coding problem for the discrete memory-less multi-level $3$-receiver broadcast channel (DM) with degraded message sets in the following subsection. In general, the setting of a broadcast channel with degraded message sets arises in video or music
broadcasting over a wireless network at varying levels of quality~\cite{elgamal}.

\begin{figure}[t]
\centering
\includegraphics[scale=0.35]{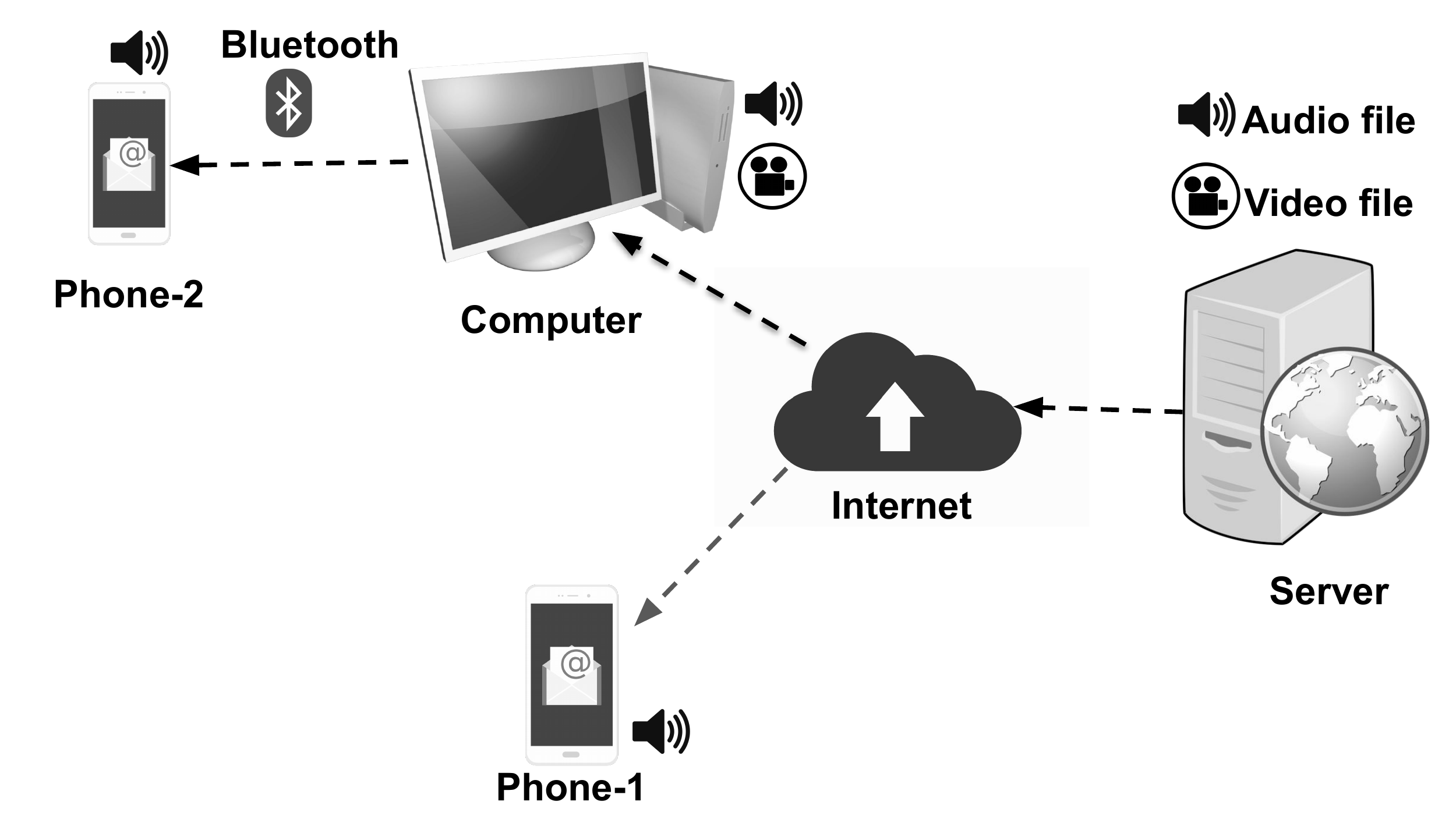}
\caption{A client-server network with $3$ clients}
\label{fig3}
\end{figure}
\subsection{Coding problem of  DM multi-level broadcast channel with degraded message sets}

The 3-receiver multi-level broadcast channel that we consider  consists of a  finite input alphabet $\mathcal{X}$ and  arbitrary  output alphabets $\mathcal{Y}_j$ for each output at the receiver-$j$ for $j \in \{1,2,3\}$. The conditional distribution of outputs at receiver-$1$ and receiver-$3$ given the input, i.e. $p_{Y_1,Y_3|X}(y_1, y_3|x)$, along with the conditional distribution of output at receiver-$2$ given the output at receiver-$1$, i.e. $p_{Y_2|Y_1}(y_2|y_1)$, are given for this broadcast channel setting, where $X$ is the input, $Y_j$ is output at receiver-$j$ for $j=1,2,3$, $x \in \mathcal{X}$ and $y_j \in \mathcal{Y}_j$ for each $j \in \{1,2,3\}$. These two conditional distributions  define this broadcast channel with three receivers since the output at the receiver-$2$ is degraded with respect to output at receiver-$1$. The broadcast channel model is shown in the Fig. \ref{fig2}.

\begin{figure}[t]
\centering
\includegraphics[scale=0.6]{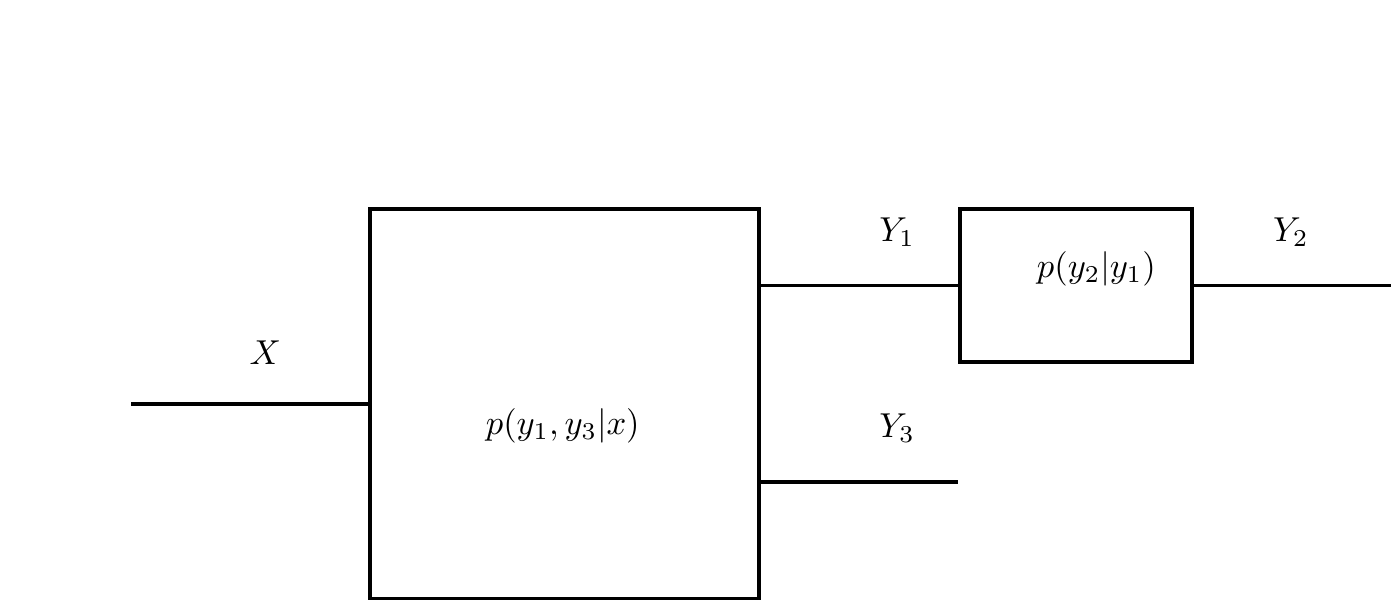}
\caption{A $3$-receiver broadcast channel model}
\label{fig2}
\end{figure}

 Now we define the coding problem whose constraint is to transmit a public message for all the receivers and a  private message intended only for receiver-$1$.

A $(2^{N{R_0}}, 2^{N{R}_1},N)$ code  consists of  
\begin{itemize}
\item a message set for public message: $\{1,2, \ldots ,2^{NR_0}\}$
\item a message set for private message of receiver-$1$: $\{1,2, \ldots ,2^{NR_1}\}$
    \item an encoder $ X^N: \{1,2, \ldots ,2^{NR_0}\} \times \{1,2, \ldots ,2^{NR_1}\} \rightarrow \mathcal{X}^n$, 
    \item a decoder at receiver-$1$ $h_1: \mathcal{Y}_1^N \rightarrow \{1,2, \ldots ,2^{NR_0}\}\times \{1,2, \ldots ,2^{NR_1}\}$,
    \item a decoder at receiver-$2$ $h_2: \mathcal{Y}_2^N \rightarrow \{1,2, \ldots ,2^{NR_0}\}$,
    \item a decoder at receiver-$3$ $h_3: \mathcal{Y}_3^N \rightarrow \{1,2, \ldots ,2^{NR_0}\}$.

\end{itemize}
where $N$ is the block length, $R_0$ is the rate of the public message and  $R_1$ is the rate of the private message. Let $M_0$ be the public message   which is chosen uniformly from the set  $\{1,2, \ldots ,2^{NR_0}\}$ and  $M_1$ be the   private message of receiver-$1$ which is chosen uniformly from the set  $\{1,2, \ldots ,2^{N R_1}\}$. Let $Y_j^{1:N}$ be the output vector at receiver-$j$ where $j \in \{1,2,3\}$. Let $P_e^{(N)} = {P}\big(  (h_1(Y_1^{1:N})  \neq (M_0,M_1)) \cup  (h_2(Y_2^{1:N})  \neq M_0) \cup (h_3(Y_3^{1:N})  \neq M_0) \big)$ be the probability of error. If there is a sequence of $(2^{NR_0},2^{NR_1}, N)$ codes, for which the $P_e^{(N)}$ goes to zero, then the rate $(R_0, R_1)$ is achieved. The closure of all such achievable rate pairs is the capacity region.
\subsection{Contribution}

In this paper, we use a polar coding strategy to achieve the rates in the capacity region for the multi-level $3$-receiver broadcast with degraded message sets without time-sharing. This represents the first time  in the literature that polar coding for 3-receiver broadcast channels without eavesdropper is considered. 

Three layered polarization results are established using auxiliary random variables that characterize the capacity region.  We do a suitable rate splitting of the private message of receiver-$1$  for the implementation of our polar coding strategy.  We use a chaining construction at two levels, one of which is within first and second layers  whereas the second level of chaining is done within the second layer.  The two-level chaining construction that we provide essentially translates into polar coding strategy the ideas of three layered superposition coding and more importantly, indirect-coding~\cite{nair} with the rate splitting of the private message.  

 The two-level chaining construction is new in the context of reliable decoding at three receivers. In particular, first level of chaining is done to recover public message by all the receivers. Second level of chaining helps to recover the split of private message reliably at receiver-$1$ while translating indirect coding of public message for receiver-$3$. In contrast, note that Marton's coding~\cite{mondelli} uses a two-level chaining construction, where first level of chaining is to align good bit-channels of the two receivers and the second level of chaining is to maintain the joint distribution of auxiliary random variables involved.

We also consider a slight variation to the problem of degraded message sets. Suppose that receiver-$1$ requires to decode only $M_1$. Then $M_1$ becomes private message to receiver-$1$ and $M_0$ is common private message to receiver-$2$ and receiver-$3$. We show that the capacity region does not enlarge by relaxing the constraint at receiver-$1$. So the same polar coding strategy achieves the capacity region of the modified problem. This is an interesting observation, as we know that for any $2$-receiver broadcast channel, superposition coding is not optimal in general, unless it is a problem with degraded message sets.
\subsection{Organization}
The paper is organized as follows. In Section \ref{np}, we introduce  some notations and recall some background results. In Section \ref{chaining}, we give our chaining  construction to achieve the rate pairs in the capacity region of the 3-receiver broadcast channel with degraded message sets and  provide the detailed decoding error analysis. In Section \ref{chaining}, we also show that the capacity of the broadcast channel remains same,  even when receiver-$1$ requires to recover only its private message. In Section \ref{conclusion}, we conclude the paper. 
\end{section}

\begin{section}{Preliminaries}\label{np}

 We denote the set $\{1,2, \ldots, n\}$ as $[n]$ where $n \in \mathcal{Z}^+$. Let $G_N$ be the conventional polar transform \cite{arikan}, represented by a binary matrix of dimension $N \times N$ where $N=2^n$, $n \in \mathcal{Z}^+$. 

Let $X$ be a binary random variable. Let the random variable pair $(X,Y)$ be distributed as $P_{X,Y}(x,y)$, then the Bhattacharya parameter is defined as
\[ Z(X|Y) = 2 \sum_{y} P_Y(y) \sqrt{P_{X|Y}(1|y)P_{X|Y}(0|y)}. \]
The following are the identities from ~\cite[Proposition ~1]{honda} which provides the relationship between entropy and Bhattacharya parameter.
\begin{equation}\label{eq:15}
(Z(X|Y))^2 \leq H(X|Y)
\end{equation}
\begin{equation}\label{eq:16}
H(X|Y) \leq \log (1 + Z(X|Y)) \leq Z(X|Y) 
\end{equation}

The \textbf{capacity region} for this multi-level $3$-receiver broadcast problem~\cite{elgamal}~\cite{nair}  is as follows:
\begin{align}
&\hspace{0.2cm}R_0 < \min \{I(W;Y_2), I(V;Y_3)\}\label{eq:1} \\
&\hspace{0.5cm}R_1 < I(X;Y_1|W)\label{eq:2}\\
&R_0 + R_1 < I(V;Y_3) + I(X;Y_1|V)\label{eq:3}
\end{align}
for some joint distribution $p(w,v)p(x|v)$ with $|\mathcal{W}| \leq |\mathcal{X}| + 4$ and $|\mathcal{V}| \leq (|\mathcal{X}| + 1)(|\mathcal{X}| + 4)$. Here $W$ and $V$ are  random variables over the alphabets $\mathcal{W}$ and $\mathcal{V}$, respectively, $Y_j$ is the output at receiver-$j$ when $X$ is input for  $j=1,2,3$.

 Let $(W_i,V_i,X_i)_{i=1}^{N}$ be the binary triplet random variable sequence that is  i.i.d. (identical and independently distributed) according to the joint distribution $p(w,v)p(x|v)$. So   $|\mathcal{X}|=|\mathcal{Y}|=|\mathcal{V}|=2$. Let  $(W,V,X)$ also be binary random triplet distributed according to $p(w,v)p(x|v)$. Let $Y_j^{1:N}$ be the received vector at receiver-$j$ when the random variable sequence $X^{1:N}$ is transmitted over the $3$-receiver discrete memoryless broadcast channel and let $Y_j$ be the output at receiver-$j$ when $X$ is input for  $j=1,2,3$. 
 
 Now we establish three-level polarization results that are going to be used in the code construction.
 
  Let $\beta  < 0.5$. Let ${(U_w)}^{1:N} = W^{1:N} G_N$, we define the following bit-channel subsets as follows where $j=1,2,3$.
\begin{align*}
&\mathcal{H}_{W} = \{ i \in [N] :  Z({(U_w)}_{i}|(U_w)^{1:(i-1)}) \geq  1 - \delta_n\} .\\
&\mathcal{L}_{W} = \{ i \in [N] :  Z((U_w)_{i}|(U_w)^{1:(i-1)}) \leq   \delta_n\}.\\
&\mathcal{H}_{W|Y_j} = \{ i \in [N] :  Z((U_w)_{i}|(U_w)^{1:(i-1)} Y_j^{1:N}) \geq  1 - \delta_n\}.\\
&\mathcal{L}_{W|Y_j} = \{ i \in [N] :  Z((U_w)_{i}|(U_w)^{1:(i-1)} Y_j^{1:N}) \leq   \delta_n\}.
\end{align*}
where $\delta_n = 2^{-N^{\beta}}$. Note that $\mathcal{L}_{W|Y_2}\subseteq \mathcal{L}_{W|Y_1}$ from {Lemma~7} in~\cite{goela} due to the degradation assumption on receiver-$2$. Then,
\begin{align*}
&\lim_{N\to\infty}  \frac{|\mathcal{H}_W|}{N}  = H(W), \hspace{1cm}\lim_{N\to\infty}  \frac{|\mathcal{L}_W|}{N}  = 1 - H(W),\\
&\lim_{N\to\infty}  \frac{|\mathcal{H}_{W|Y_j}|}{N}  = H(W|Y_j), \lim_{N\to\infty}  \frac{|\mathcal{L}_{W|Y_j}|}{N}  = 1 - H(W|Y_j).
\end{align*}
Let ${(U_v)}^{1:N} = V^{1:N}G_N $. We now define  bit-channel subsets $\mathcal{H}_{V|W}$ and  $\mathcal{L}_{V|W}$ based on the Bhattacharyya parameter $Z({(U_v)}_{i}|(U_v)^{1:(i-1)}W^{1:N})$ as we did above. Similarly we define the $\mathcal{H}_{V|W Y_j}$ and  $\mathcal{L}_{V|W Y_j}$ based on the value of Bhattacharyya parameter $Z({(U_v)}_{i}|(U_v)^{1:(i-1)}W^{1:N}Y_j^{1:N})$ for $j=1,3$. Then,
\begin{align*}
&\lim_{N\to\infty}  \frac{|\mathcal{H}_{V|W}|}{N}  = H(V|W) , \lim_{N\to\infty}  \frac{|\mathcal{L}_{V|W}|}{N}  = 1 - H(V|W),\\
&\lim_{N\to\infty}  \frac{|\mathcal{H}_{V|WY_j}|}{N}  = H(V|WY_j),\\&\lim_{N\to\infty}  \frac{|\mathcal{L}_{V|WY_j}|}{N}  = 1 - H(V|WY_j).
\end{align*}
Let ${(U_x)}^{1:N} = X^{1:N}G_N $. We define the  bit-channel subsets $\mathcal{H}_{X|V}$, $\mathcal{L}_{X|V}$ and also $\mathcal{H}_{X|V Y_1}$, $\mathcal{L}_{X|V Y_1}$ based on the values of Bhattacharyya parameters $Z((U_x)_{i}|(U_x)^{1:(i-1)}V^{1:N})$ and $Z((U_x)_{i}|(U_x)^{1:(i-1)}V^{1:N} Y_1^{1:N})$, respectively, as we did above. Then,
\begin{align*}
&\lim_{N\to\infty}  \frac{|\mathcal{H}_{X|V}|}{N}  = H(X|V) ,\lim_{N\to\infty}  \frac{|\mathcal{L}_{X|V}|}{N}  = 1 - H(X|V).\\
&\lim_{N\to\infty}  \frac{1}{N} |\mathcal{H}_{X|VY_1}| = H(X|V Y_1),\\
&\lim_{N\to\infty}  \frac{1}{N} |\mathcal{L}_{X|VY_1}| = 1 - H(X|VY_1).
\end{align*}

Under this probability distribution  of $(W^{1:N}, V^{1:N}, X^{1:N})$, we denote
  $\mathbb{P}((U_w)^{1:N} = (u_w)^{1:N})$
by $P_{(U_w)^{1:N}}((u_w)^{1:N})$ and similarly we denote $\mathbb{P}((U_v)_i = (u_v)_i |W^{1:N} (U_v)^{1:i-1}Y_1^{1:N} = w^{1:N} u_v^{1:i-1}y_1^{1:N}  )$ by $ P_{(U_v)_i|W^{1:N} (U_v)^{1:i-1}Y_1^{1:N}}((u_v)_i|w^{1:N}(u_v)^{1:i-1}y_1^{1:N})$.

Define $I_j^w = \mathcal{L}_{W|Y_j} \cap \mathcal{H}_{W}$, $I_j^v = \mathcal{L}_{V|WY_j} \cap \mathcal{H}_{V|W}$ and $I_j^x = \mathcal{L}_{X|VY_1} \cap \mathcal{H}_{X|V}$. Note that $\lim_{N \to \infty}\frac{|I_j^w|}{N} = I(W;Y_j)$, $\lim_{N \to \infty}\frac{|I_j^v|}{N} = I(V;Y_j|W)$ and $\lim_{N \to \infty}\frac{|I_j^x|}{N} = I(X;Y_j|V)$. We refer to $I_j^w$, $I_j^v$ and $I_j^x$ as information bit-channels of receiver-$j$ in $(U_w)^{1:N}$,  $(U_v)^{1:N}$ and $(U_x)^{1:N}$ respectively for $j=1,2,3$.

Define $F_j^w = \mathcal{H}_W - I_j^w $, $F_j^v = \mathcal{H}_{V|W} - I_j^v $ and $F_j^x = \mathcal{H}_{X|V} - I_j^x $. We refer to $F_j^w$, $F_j^v$ and $F_j^x$ as frozen bit-channels of receiver-$j$ in $(U_w)^{1:N}$,  $(U_v)^{1:N}$ and $(U_x)^{1:N}$ respectively for $j=1,2,3$. 

Define $R^w = (\mathcal{H}_W \cup \mathcal{L}_W)^c$, $R^v = (\mathcal{H}_{V|W} \cup \mathcal{L}_{V|W})^c$ and $R^x = (\mathcal{H}_{X|V} \cup \mathcal{L}_{X|V})^c$. We refer to $R^w$, $R^v$ and $R^x$ as not-completely polarized bit-channels in $(U_w)^{1:N}$,  $(U_v)^{1:N}$ and $(U_x)^{1:N}$ respectively.

We denote the subvector of $U^{1:N}$ corresponding to the bit-channel set $\mathcal{A} \subset [N]$  by $U^{\mathcal{A}}$. Let $P$ and $Q$ be any two distributions on a discrete arbitrary alphabet $\mathcal{Z}$. We denote the total variation distance between the two distributions $P$ and $Q$ as $|| P-Q ||$. Therefore $|| P-Q || = \sum_{z \in \mathcal{Z}} \frac{1}{2}|P(z) - Q(z)| = \sum_{z: P(z) > Q(z)} P(z) - Q(z).$ We denote the KL-divergence between two distributions $P$ and $Q$ as $D( P||Q)$.
\end{section}
\begin{section}{Polar  coding for the DM Multi-level $3$-receiver broadcast channel}\label{chaining}
In this section, we are going to discuss the polar coding scheme for achieving the capacity region of the DM multi-level $3$-receiver broadcast channel with degraded message sets. To achieve the capacity region, we need to achieve the rate pairs that satisfy equations (\ref{eq:1}), (\ref{eq:2}) and (\ref{eq:3}) for all joint distributions on random variables over the alphabets of the required size  mentioned in the definition of the capacity region. We consider the case when $|\mathcal{X}|= |\mathcal{V}|=|\mathcal{W}|=2$ to describe the polar coding scheme. The fundamental idea of the polar coding strategy which we present is applicable even when the alphabets $|\mathcal{X}|, |\mathcal{W}|$ or $|\mathcal{V}|$ are of higher size. In \cite{sasoglu1} \cite{sasoglu2}, polarization for the alphabets of higher size is discussed.

\subsection{Typical set coding}

Before we go into our polar coding construction, we briefly discuss the achievability of the rate pairs in the capacity region using random coding approach by typical sets \cite[p.~200]{elgamal}. Three layered superposition coding with a rate splitting of the private message and  indirect coding of public message at receiver-3 are used in the scheme. Let $N$ be the block length. Let $R_1 = R_{11} +R_{12}$ be the rate split of the private message. We first generate $2^{NR_0}$-$w^N$ sequences, whose components are i.i.d. according to the distribution $p(w)$, independently for the public message. Then we use superposition coding to generate $2^{NR_{11}}$-$v^N$  sequences, whose components are independent according to conditional distribution $p(v|w)$ given each $w^N$ sequence, independently  for the part of private message. We again use  superposition coding to generate $2^{NR_{12}}$-$x^N$  sequences,  whose components are independent according to conditional distribution $p(x|v)$ given each $v^N$ sequence, independently for the other part of private message. For each public message and private message pair, their corresponding $x^N$ sequence gets transmitted as a codeword. Receiver-$1$ recovers the unique public message and private message pair whose $(w^N, v^N, x^N)$ is jointly typical with received sequence at the receiver. Receiver-$2$ recovers the unique public message whose $w^N$ is jointly typical with received sequence at the receiver.  Instead of recovering public message like how receiver-$2$ does, receiver-$3$ recovers the unique public message whose $w^N$ sequence and at-least one of its $v^N$ sequence in second layer  is jointly typical with received sequence at the receiver, which is referred to as indirect decoding method. If $R_0$, $R_1$, $R_{11}$, $R_{12}$ satisfy the following: 
 \begin{align}
&\hspace{0.5cm}R_0 < I(W; Y_2)\label{eq:8} \\
&\hspace{0.5cm}R_{12} < I(X;Y_1|V)\label{eq:9}\\
&\hspace{0.2cm}R_{11} + R_{12} < I(X;Y_1|W)\label{eq:10}\\
&\hspace{0.1cm}R_0 + R_{11} + R_{12} < I(X;Y_1)\label{eq:11}\\
&\hspace{0.3cm}R_0 + R_{11} < I(V;Y_3) , \label{eq:12}
\end{align}
then reliable recovery of the intended messages at each of the receivers is ensured. After eliminating variables $R_{11}$ and $R_{12}$ by Fourier-Motzkin procedure~\cite{elgamal} by substituting $R_1 = R_{11} + R_{12}$, we get the region described by equations $(\ref{eq:1})$, $(\ref{eq:2})$ and $(\ref{eq:3})$ that defines the capacity region.

The intuition behind the rate splitting is that if we want to achieve a private message rate satisfying $R_1 > I (X; Y_1|V)$ and $R_1 < I(X;Y_1|W)$, then we rate split $R_1$ into $R_{11}$ and $R_{12}$ such that $R_{12} < I (X;Y_1|V)$. As we recover public message indirectly using $v^N$ sequences at receiver-$3$, the sum of public message rate $R_0$ and $R_{11}$ should be less than $I(V;Y_3)$. So, if we make $R_{11}$ small while rate splitting, then it can be noticed that the public message rate can be improved, provided the reliability constraint at receiver-$2$, $R_0 < I(W;Y_2)$, is loose.

\subsection{Rate splitting of the private message for polar coding}

Notice that a point in  the region satisfied by equations (\ref{eq:8}), (\ref{eq:9}),  (\ref{eq:10}), (\ref{eq:11}) and (\ref{eq:12}) does not always satisfy the constraint $R_{11} < I(V; Y_1|W)$. We impose the new additional constraint  $R_{11} < I(V;Y_1|W)$ for the rate split in the implementation of our polar coding strategy through following lemma.

\begin{lemma}\label{lemma:1}
For any rate pair $(R_0, R_1)$ that satisfies equations $(\ref{eq:1})$ $(\ref{eq:2})$ and $(\ref{eq:3})$ and for a particular joint distribution $p(w,v)p(x|v)$ on $(W, V, X)$, there exist rates $R_{11}$ and $R_{12}$ such that $R_1 = R_{11} + R_{12}$ (rate split of $R_1$)  and following three identities hold.
\begin{center}
$R_{11} <  I(V; Y_1|W)$\\$  R_{12} <  I(X; Y_1|V) $\\
$R_0 + R_{11} < I(V; Y_3)$
\end{center}
\end{lemma} 
\noindent
\textbf{Proof:}\\
It is easy to find the split for $R_1$ such that the first two identities hold since $I(V; Y_1|W)+I(X; Y_1|V) =I(X;Y_1|W)$ ($W\rightarrow V \rightarrow X\rightarrow Y_1$ is chain). Let $R_{11}'$ and $R_{12}'$ be such a rate split for $R_1$. Suppose that the third identity does not hold for the split $R_1 = R_{11}'+ R_{12}'$. That means $R_0 + R_{11}' \geq I(V; Y_3)$. Say that $R_0 + R_{11}' = I(V; Y_3)+ \delta$ for some $\delta \geq 0$. On the other hand we have
$R_0 + R_{11}' + R_{12}' < I(V; Y_3) + I(X;Y_1|V)$. So we should have $R_{12}' < I(X; Y_1|V) - \delta$. Say that $R_{12}' = I(X; Y_1|V) - \delta_1$. Clearly $\delta_1 > \delta$.

Note that $R_{11}' > \delta$, since $R_0 <I(V; Y_3)$. Choose $R_{11} = R_{11}' - \delta^+$ and $R_{12} =  R_{12}' + \delta^+$ where $\min\{R_{11}', \delta_1\} > \delta^+ > \delta$. Clearly, $R_{11}$ and $R_{12}$ is a split of $R_1$ that satisfies the required three identities. Hence the claim of the lemma is shown. \qed

In our polar coding strategy, the private message bits for receiver-$1$ are given in bits $I_1^v$ and $I_1^x$ that are corresponding to $V^N$ vectors and $X^N$ vectors, which are involved in the chaining construction we provide, respectively. The rate split in  Lemma \ref{lemma:1} allows us to associate the private message bits encoded in $I_1^v$  and $I_1^x$ to split rates of the private message $R_{11}$ and $R_{12}$, respectively.  We also involve the bits corresponding to $R_{11}$, which are private message bits encoded in $I_1^v$, in the chaining procedure to translate the indirect coding method at receiver-3 into polar coding.  We also use the degradation condition of receiver-$2$ in our code construction. Now we provide our code construction in the following subsection.

\subsection{Code construction}

We give a polar coding strategy for each of the following possible cases for the rate pair $(R_0,R_1)$.
\begin{itemize}
    \item $ R_0 \geq I(W;Y_3)$
    \item $ R_0 < I(W;Y_3)$
\end{itemize}

We consider $k$ polar blocks of size $N$ large enough so that the polarization happens. We propose a chaining construction with these $k$ polar blocks for the rate pair $(R_0,R_1)$ by using the rate split given by the Lemma  \ref{lemma:1}. 

While encoding each polar block, we first construct $(U_w)^{1:N}$ and compute $W^{1:N} =(U_w)^{1:N} G_N $. We next construct $V^{1:N}=(U_v)^{1:N}G_N$ given ${W^{1:N}}$ and apply polar transform to obtain $V^{1:N}$. Lastly, we construct $(U_x)^{1:N}$ 
  given $V^{1:N}$ and apply polar transform to obtain $X^{1:N}$ (codeword). This encoding method ensures that the average distribution of $(W_i, V_i, X_i )_{i=1}^N$ is close in total variation distance to the distribution which is induced when $(W_i, V_i, X_i )_{i=1}^N$ is  i.i.d.  according to $p(w)p(v|w)p(x|v)$. The total variation distance becomes $O(2^{-N^{\beta'}})$ where $\beta' < \beta < 0.5$.

  We first give the construction for the case where $R_0 \geq I(W;Y_3)$. This is the case where we translate the indirect coding into polar coding strategy. We assume $NR_{11} > |I_1^v \cap I_3^v|$  to demonstrate the code construction. The construction we give under this assumption gives the general idea of the chaining construction which can easily be extended to the case where this assumption does not hold. 
  
  Note that public message bits have to be recovered at all the receivers. If we give $NR_0$ public message bits in $I_2^w$, receiver-$2$ and receiver-$1$ (due to degradation condition) can recover these bits. But receiver-$3$ may not be able to decode in that case.On the other hand we can recover these bits at receiver-$3$, if we place these bits into  $I_3^w$ and  remaining $NR_0-|I_3^w|$ bits in  $I_3^v$, as $NR_0 > |I_3^w|$. In this case, In this case, receiver-$1$ and receiver-$2$  may not be able to decode. We do a chaining, to resolve the alignment of the bit-channel set in 
  $I_2^w$  with  bit-channels sets in $I_3^w$ and $I_3^v$ to allocate the public message bits for reliable recovery at all the receivers.
  
  Since we are assigning a portion of public message bits in  $(U_v)^{1:N}$ vectors  for receiver-$3$, we need to recover $(U_v)^{1:N}$ vectors at receiver-$3$. But we also use $(U_v)^{1:N}$ vectors  for encoding private message bits corresponding to the rate $R_{11}$.  If we give these private message bits in $I_1^v$, receiver-$3$ cannot recover these bits, which blocks receiver-$3$ from recovering $(U_v)^{1:N}$ vectors  for decoding the portion of intended public message bits. Here is where we need to do a second level of chaining for aligning bit-channel set in $I_3^v$ with bit-channel set in $I_1^v$ where we provide  private message bits corresponding to $R_{11}$. This summarizes the main idea behind the construction that translates indirect coding at receiver-$3$.
  
  Fig. \ref{fig4} shows how we fill $(U_w)^{1:N}$, $(U_v)^{1:N}$ and $(U_x)^{1:N}$ vectors when $k=3$ allocating public and private message bits.   The links between vectors in Fig. \ref{fig4} indicate  the copying of  bits between bit-channel sets of successive blocks.
Now we provide detailed steps in encoding and decoding methods in the two-level chaining construction for this case, $R_0 \geq I(W;Y_3)$. \\
\textbf{Encoding:}
\begin{itemize}
\item Encoding $(k-1) NR_0 + |I_3^w \cap I_2^w|$ bits of the public message, first level of chaining:
\begin{itemize}
\item
We first place $|I_3^w \cap I_2^w|$  bits in $(U_w)^{I_3^w \cap I_2^w}$ for all the blocks $t=1:k$.  Note that $NR_0$ is the sum of  $|I_3^w \cap I_2^w| + |I_3^w \cap F_2^w| + (NR_0 - |I_3^w|)$.
\item 
We place $|I_3^w \cap F_2^w|$ bits in $(U_w)^{I_3^w \cap F_2^w}$ and $NR_0 - |I_3^w|$ in $(U_v)^{I_{31}^v}$ for the blocks $t=1:k-1$  where $I_3^v$ is partitioned as disjoint union $I_{31}^v \cup I_{32}^v$,  $|I_{31}^v| = NR_0-|I_3^w|$. Note that $NR_0 + NR_{11}$ < $|I^w_3| + |F_1^v \cap I_3^v| + |I_1^v \cap I_3^v|$ due to Lemma \ref{lemma:1}. As we assumed the case where $NR_{11} > |I_1^v \cap I_3^v|$, it can be deduced that  $I_{31} \subset I_{3}^v\cap F_{1}^v$. 
\item 
We copy bits in $(U_w)^{I_3^w \cap F_2^w}$ and $(U_v)^{I_{31}^v}$ of block $t$ to $(U_w)^{B_{w1}}$ of block $t+1$ for $t=1:k-1$ where $I_2^w \cap F_3^w$ is partitioned as disjoint union $B_{w1} \cup B_{w2}$ and $|B_{w1}| = NR_0 - |I_3^w \cap I_2^w|$. 
\end{itemize}
\item 
Encoding $(k-1) NR_{11} + |I_1^v \cap I_3^v|$ bits of the private message for receiver $1$, second level of chaining: 
\begin{itemize}
\item We first place $|I_3^v \cap I_2^v|$  private message bits in $(U_v)^{I_3^v \cap I_2^v}$ for all the blocks $t=1:k$. 
\item 
We place $NR_{11}- |I_3^v \cap I_2^v|$ bits in $(U_v)^{I_{321}^v} $ for the blocks $t=1:k-1$  where  $I_{32}^v$ is partitioned as disjoint union $I_{321}^v \cup I_{322}^v \cup (I_1^v \cap I_3^v)$,  and $ |I_{321}^v|=NR_{11}- |I_1^v \cap I_3^v|$.  Note that $NR_{11} < \text{min}\{|I_{32}^v|,|I_{1}^v|\}$ due to Lemma \ref{lemma:1}. 
\item We copy the bits in $(U_w)^{I_{321}^v}$ of block $t$ to $(U_v)^{I_{11}^v}$ of block $t+1$ for $t=1:k-1$, where $ (I_1^v \cap F_3^v)$ is partitioned as the disjoint union $I_{11}^v \cup I_{12}^v$ and  $|I_{11}^v| = |I_{321}^v|$. 
\end{itemize}
\item 
Encoding $k NR_{12}$ bits of the private message for receiver-$1$: We place $NR_{12}$ bits in $(U_x)^{I_{1}^x}$ for all these blocks $t=1:k$. Note that $NR_{12} < |I_{1}^x|$ due to Lemma \ref{lemma:1}. We do not involve this portion of the private message bits in the chaining.  

\item 
We place randomly chosen frozen bits with i.i.d. uniform distribution in $(U_w)^{B_{w1}}$, $(U_v)^{I_{11}^v}$ for the block $t=1$. We place randomly chosen frozen bits with i.i.d. uniform distribution in  $(U_w)^{(I_3^w \cap F_2^w)}$, $(U_v)^{I_{31}^v}$, $(U_v)^{I_{321}^v}$ for the block $t=k$.   We place randomly chosen bits with i.i.d. uniform distribution in the remaining positions of $(U_w)^{\mathcal{H}_{W}}$, $(U_v)^{\mathcal{H}_{V|W}} $ and $(U_x){^{\mathcal{H}_{X|V}}}$, which are not filled by private or public message bits, in all the $k$ blocks. We share these remaining bits that are in  $(U_w)^{F_j}$, $(U_v)^{F_j}$ and $(U_x)^{F_j}$ of each block with the receiver-$j$ for $j=1,2,3$, in all the $k$ blocks.
\item We have constructed  $(U_w)^{\mathcal{H}_{W}}$, $(U_v)^{\mathcal{H}_{V|W}}, (U_x)^{\mathcal{H}_{X|V}}$ for all the $k$ blocks. Now we encode other positions in $(U_w)^{1:N}$, $(U_v)^{1:N}, (U_x)^{1:N}$ as we do for single asymmetric channel case~\cite{honda}, \cite{karthik} for all the blocks $t=1:k$. 
\item We use the  following decision rule for encoding  $(U_w)^{\mathcal{L}_{W}}$.
\begin{equation*}
 (U_w)_i= \text{argmax}_{x \in \{0,1\}} P_{(U_w)_i|(U_w)^{1:{i-1}})}(x|(U_w)^{1:{i-1}}).
 \end{equation*}
 For $i \in \mathcal{L}_W$, the induced conditional distribution $\delta_i^w((u_w)_i|(u_w)^{1:{i-1}})$ on $(U_w)_i$ given $(U_w)^{1:i-1}$ satisfies ${\delta_i^w((u_w)_i| (u_w)^{1:i-1}) = 1}$ and $\delta_i((u_w)_i+1| (u_w)^{1:i-1
}) = 0$ where
\begin{equation*}
(u_w)_i = \text{argmax}_{x \in \{0,1\}}
P_{(U_w)_i|(U_w)^{1:{i-1}}}(x|(u_w)^{1:{i-1}}) .
\end{equation*}
 \item We use either randomly chosen boolean functions, which are shared with all the receivers, or common randomness~\cite{honda}~\cite{karthik} for encoding bit-channels in  $(U_w)^{R_w}$  to maintain the conditional distribution $P_{(U_w)_i|(U_w)^{1:{i-1}}}$ on an average over the random ensemble. We now compute $W^{1:N} = (U_w)^{1:N}G_N$.
 \item 
We use the decision rule below  for encoding $(U_v)^{\mathcal{L}_{V|W}}$.
\begin{align*}
 (U_v)_i  &= \text{argmax}_{x \in \{0,1\}} \\& \text{\hspace{1cm}}P_{(U_v)_i|W^{1:N}(U_v)^{1:{i-1}}}(x|W^{1:N} (U_v)^{1:{i-1}}).
\end{align*}
 For $i \in \mathcal{L}_{V|W}$, the induced conditional distribution $\delta_i^v((u_v)_i|w^{1:n} (u_v)^{1:{i-1}} )$ on $(U_v)_i$ given $W^{1:N}(U_v)^{1:i-1}$ satisfies ${\delta_i^v((u_v)_i| w^{1:N}(u_v)^{1:i-1}) = 1}$ and $\delta_i^v((u_v)_i+1| (u_v)^{1:i-1}) = 0$ where
\begin{align*}
(u_v)_i &= \text{argmax}_{x \in \{0,1\}}\\& \text{\hspace{1cm}}
P_{(U_v)_i|W^{1:N}(U_v)^{1:{i-1}}}(x|w^{1:N}(u_v)^{1:{i-1}}) .
\end{align*}
 \item We use either randomly choosen boolean functions, which are shared with all the receivers, or common randomness  for encoding  bit-channels in $(U_v)^{R_v}$ to maintain the conditional distribution $P_{(U_v)_i|W^{1:N}(U_v)^{1:{i-1}} }$.  Now we compute $V^{1:N} = (U_v)^{1:N}G_N$.
 \item 
We use the decision rule below  for encoding $(U_x)^{\mathcal{L}_{X|V}}$.
\begin{align*}
 (U_x)_i  &= \text{argmax}_{x \in \{0,1\}} \\& \text{\hspace{1cm}}P_{(U_x)_i|V^{1:N}(U_x)^{1:{i-1}}}(x|V^{1:N} (U_x)^{1:{i-1}}).
\end{align*}
 For $i \in \mathcal{L}_{X|V}$, the induced conditional distribution $\delta_i^x((u_x)_i|v^{1:n} (u_x)^{1:{i-1}} )$ on $(U_x)_i$ given $V^{1:N}(U_x)^{1:i-1}$ satisfies ${\delta_i^x((u_x)_i| v^{1:N}(u_x)^{1:i-1}) = 1}$ and $\delta_i^x((u_x)_i+1| (u_x)^{1:i-1}) = 0$ where
\begin{align*}
(u_x)_i &= \text{argmax}_{x \in \{0,1\}}\\& \text{\hspace{1cm}}
P_{(U_x)_i|V^{1:N}(U_x)^{1:{i-1}}}(x|v^{1:N}(u_x)^{1:{i-1}}) .
\end{align*}
\item We use either randomly chosen boolean functions, which are shared with all the receivers, or common randomness for encoding  bit-channels in $(U_x)^{R_x}$ to maintain the conditional distribution $P_{(U_x)_i|V^{1:N} (U_x)^{1:{i-1}} }$.
Now we compute $X^{1:N} = (U_x)^{1:N}G_N$.
\item We transmit $X^{1:N}$ for all $k$ blocks.
\end{itemize}
\begin{figure}[t]
\centering
\includegraphics[scale=0.45]{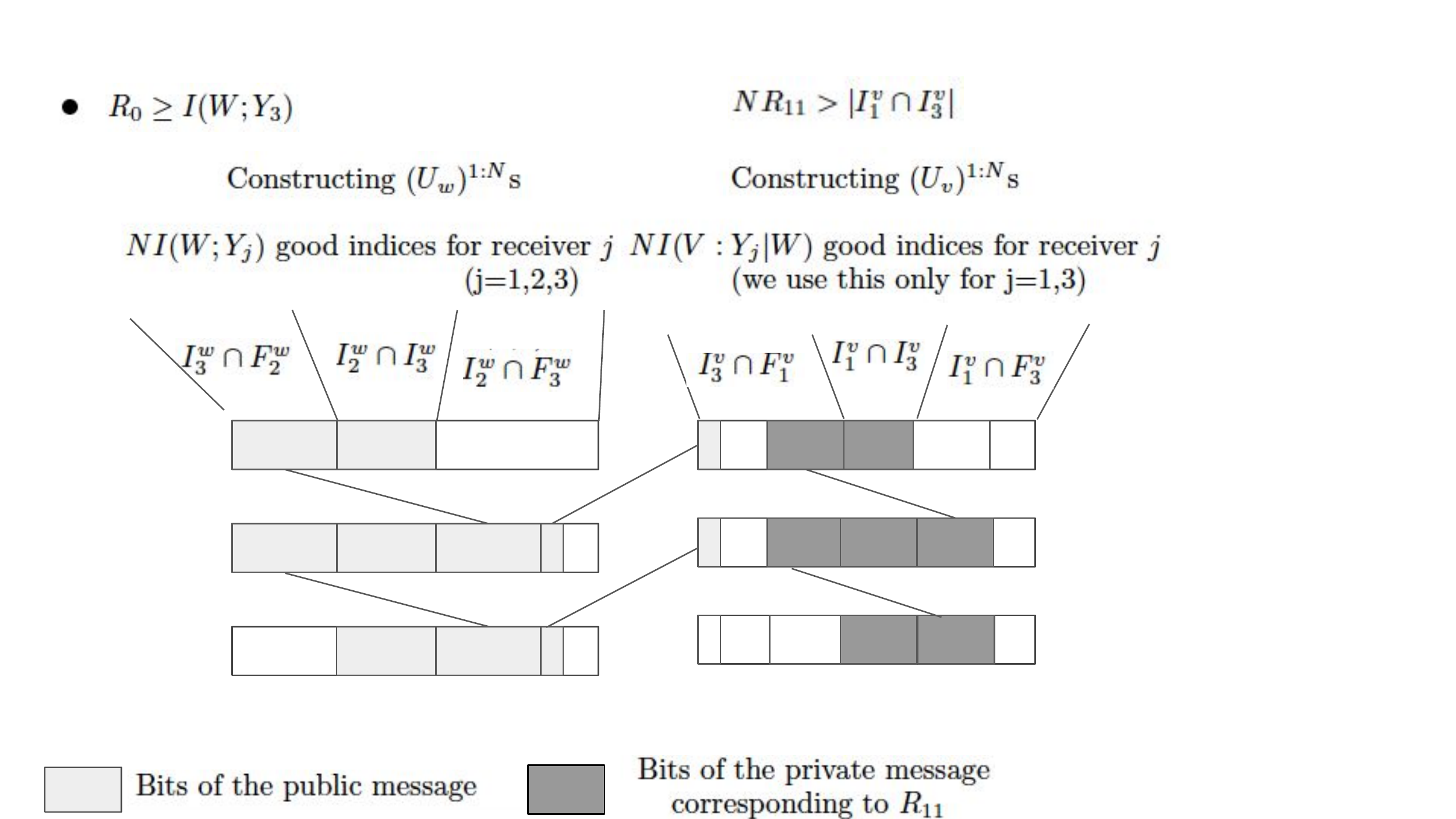}
\includegraphics[scale=0.45]{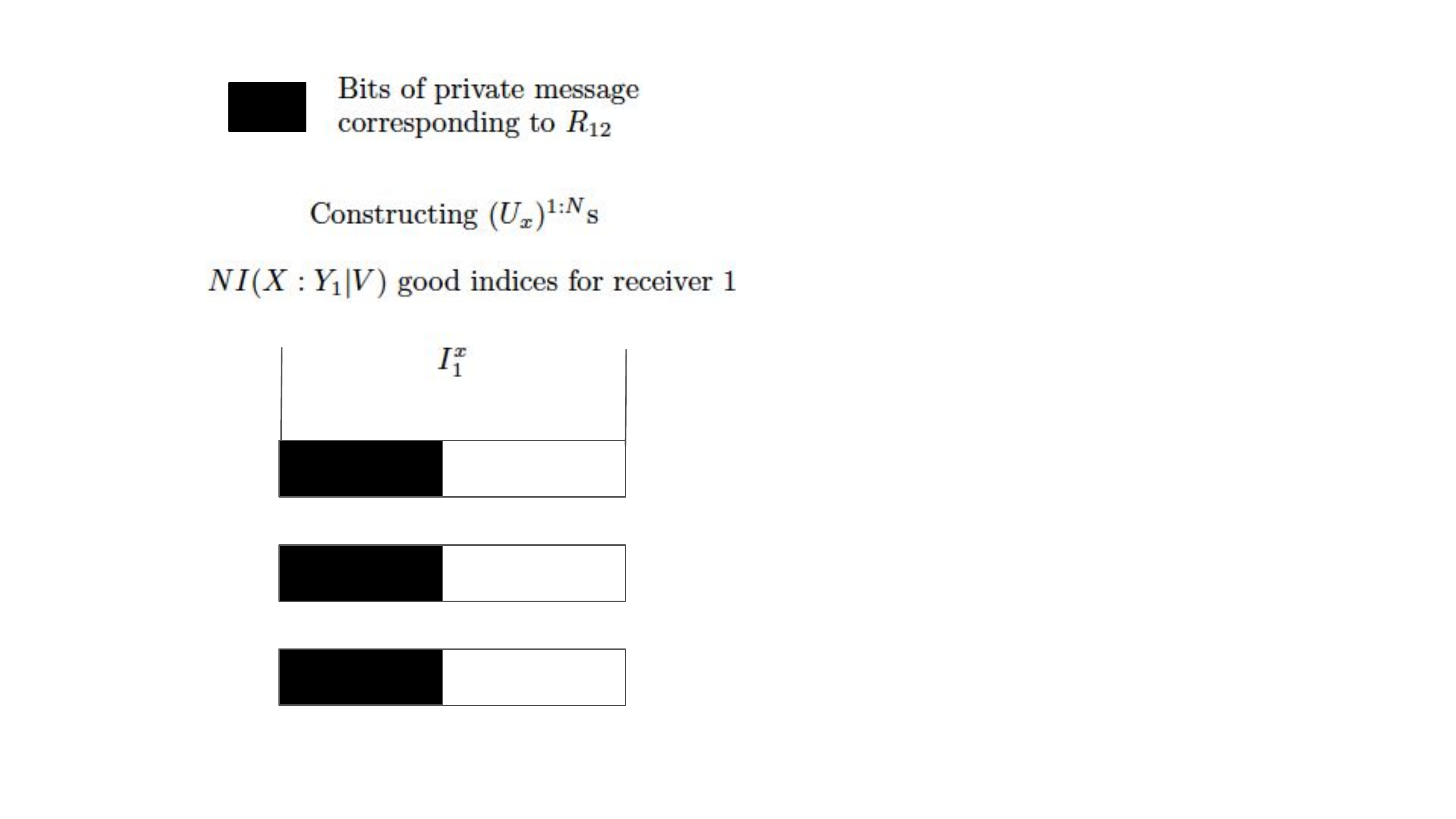}
\vspace{-1.0cm}
\caption{Private and public message bits allocation in $(U_w)^{1:N}$, $(U_v)^{1:N}$ and $(U_x)^{1:N}$ vectors when $k=3$ }
\label{fig4}
\end{figure}
\textbf{Rate of scheme:} We encoded $(k-1) \cdot NR_0 +  |I_2^w \cap I_3^w|$ public message bits for $k$ blocks. We encoded $(k-1) \cdot NR_{11} + k \cdot NR_{12} +  |I_1^v \cap I_3^v|$ private message bits for $k$ blocks. Hence the we achieve the rate pair $(\frac{(k-1) \cdot NR_0 +   |I_2^w \cap I_3^w|}{k \cdot N}, \frac{(k-1) \cdot NR_{11} + k \cdot NR_{12} +  |I_1^v \cap I_3^v|}{k \cdot N})$, which approaches the pair $(R_0, R_1)$ as $k$ goes infinity.  Now we provide the decoding method for the case $R_0 \geq I(W;Y_3)$.\\

\textbf{Decoding, using $Y_j^{1:N}$ at receiver-$j$ for all $k$ blocks:}
\begin{itemize}
\item   The following steps $1)-4)$ give the decoding procedure at receiver-$3$.  We decode both $(U_w)^{1:N}$s and $(U_v)^{1:N}$s of the blocks to recover the public message bits at receiver-$3$. 
\begin{enumerate}
 \item
 Set $t=1$. We decode $(U_w)^{1:N}$ and $(U_v)^{1:N}$  by successive cancellation  for the block $t$. $NR_0$ bits  $(U_w)^{I_3^w}$ and $(U_v)^{I_{31}^v}$ of the public message will be recovered in this step. $NR_{11}$  bits in $(U_v)^{I_{32}^v}$ of the private message of receiver-$1$ will also be  recovered. 
\item   We decode $(U_w)^{1:N}$  followed by $(U_v)^{1:N}$ by successive cancellation for the block $t+1$. The bits  $(U_w)^{I_3^w \cap F_2^v}$, $(U_v)^{I_{31}^v}$ and $(U_v)^{I_{321}^v}$ recovered for block $t$  give bits in  $(U_w)^{B_{w1}}$ and $(U_v)^{I_{11}^v}$ during the successive cancellation decoding of block $t+1$. $NR_0$ bits  $(U_w)^{I_3^w}$ and $(U_v)^{I_{31}^v}$ of the public message will be recovered in this step. $NR_{11}$ bits in  $(U_v)^{I_{32}^v}$ of the private message of receiver-$1$  will also be  recovered.  Increase $t$ by $1$.
\item  Repeat step (2) until $t=k-1$.
\item  We decode $(U_w)^{1:N}$ and $(U_v)^{1:N}$ in successive cancellation style for block $k$. The bits  $(U_w)^{I_3^w \cap F_2^v}$, $(U_v)^{I_{31}^v}$ and $(U_v)^{I_{321}^v}$ and  recovered for block $k-1$  give bits in  $(U_w)^{B_{w1}}$ and $(U_v)^{I_{11}^v}$ during the successive cancellation decoding of block $k$. The bits  $(U_w)^{I_3^w \cap I_2^w}$ of the public message will be recovered in this step.
\end{enumerate}
\item The following steps (1)-(4) give the decoding procedure at receiver-$1$. The content in parentheses-\big(\big) is ignored when decoding at receiver-$2$. We decode all $(U_w)^{1:N}$s, $(U_v)^{1:N}$s and of $(U_x)^{1:N}$s of all the blocks to recover the public message bits and private message bits at receiver-$1$. We only decode $(U_w)^{1:N}$s of all the blocks to recover public message  bits at receiver-$2$.

\begin{enumerate}
\item   Set $t=k$. We decode $ (U_w)^{1:N}$ \big(,$(U_v)^{1:N}$ and $(U_x)^{1:N}$\big) by successive cancellation  for block $t$. $NR_0$ bits $(U_w)^{I_3^w \cap I_2^w}$ and $(U_w)^{B_{w1}}$  of the public message will be recovered  for block $t$. \big($NR_{11}$ bits in $(U_v)^{I_{11}^v \cup (I_{1}^v \cap I_3^v)}$ of the private message of receiver-1 will also be recovered. $NR_{12}$  bits in $(U_x)^{I_1^x}$ of the private message of receiver-$1$  will also be recovered.\big)
\item We decode  $(U_w)^{1:N}$\big(,$(U_v)^{1:N}$ and $(U_x)^{1:N}$\big) by successive cancellation for block $t-1$. The bits  $(U_w)^{B_{w1}}$, \big($(U_v)^{I_{11}^v}$\big) recovered for block $t$ give bits in $(U_w)^{I_3^w \cap F_2^w}$ \big(, $(U_v)^{I_{31}^v}$ and  $(U_v)^{I_{321}^v}$\big) during the successive cancellation decoding of block $t-1$.  $NR_0$ bits   $(U_w)^{I_3^w \cap I_2^w}$ and  $(U_w)^{B_{w1}}$ of the public message will be recovered.\big($NR_{11}$ bits in $(U_v)^{I_{11}^v \cup (I_{1}^v \cap I_3^v)}$ of the private message of receiver-1 will also be recovered. $NR_{12}$  bits in $(U_x)^{I_1^x}$ of the private message of receiver-$1$  will also be recovered.\big) Decrease $t$ by $1$.
\item Repeat step (2) until $t=2$.
\item  We decode $ (U_w)^{1:N}$ \big(,$(U_v)^{1:N}$ and $(U_x)^{1:N}$\big) by successive cancellation for block $1$. The bits  $(U_w)^{B_{w1}}$ \big(, $(U_v)^{I_{11}^v}$\big) recovered for block $2$ give bits in $(U_w)^{I_3^w \cap F_2^w}$ \big(, $(U_v)^{I_{31}^v}$ and  $(U_v)^{I_{321}^v}$\big) during the successive cancellation decoding of block $1$. The bits $(U_w)^{I_3^w \cap I_2^w}$ of the public message will be recovered. \big(The bits  $(U_v)^{I_{1}^v\cap I_3^v }$ of the private message of receiver-1  will also be recovered. $NR_{12}$  bits in $(U_x)^{I_1^x}$ of the private message of receiver-$1$  will also be recovered.\big)
\end{enumerate}
\item During the successive cancellation decoding, we recover the needed bits in $(U_w)^{\mathcal{L}_{W|Y}}$, $(U_x)^{\mathcal{L}_{V|W}}$ and  $(U_v)^{\mathcal{L}_{X|V}}$ at each receiver by an appropriate decision/arg-max rule.
\item We use the  following decision rule for decoding  $(U_w)^{\mathcal{L}_{W}}$ at receiver-$j=1,2,3$.
\begin{align*}
 (U_w)_i&= \text{argmax}_{x \in \{0,1\}}\\& \text{\hspace{1cm}} P_{(U_w)_i|(U_v)^{1:{i-1}} Y_j^{1:N}}(x|(U_w)^{1:{i-1}}Y_j^{1:N}).
 \end{align*}
We use the following decision rule for decoding $(U_v)^{\mathcal{L}_{V|W}}$ at reciever-$j=1,3$.
\begin{align*}
 (U_v)_i  &= \text{argmax}_{x \in \{0,1\}} \\& \text{\hspace{0.23cm}}P_{(U_v)_i|W^{1:N}(U_v)^{1:{i-1}}Y_j^{1:N}}(x|W^{1:N} (U_v)^{1:{i-1}}Y_j^{1:N}).
\end{align*}
We use the following decision rule below  for decoding $(U_x)^{\mathcal{L}_{X|V}}$ at reciever-$1$.
\begin{align*}
 (U_x)_i  &= \text{argmax}_{x \in \{0,1\}} \\& \text{\hspace{0.23cm}}P_{(U_x)_i|V^{1:N}(U_x)^{1:{i-1}}Y_1^{1:N}}(x|V^{1:N} (U_x)^{1:{i-1}}Y_1^{1:N}).
\end{align*}
\item The remaining  bits could be either the bits in frozen positions which are available at the corresponding receiver or the bits in  $(U_w)^{R_w}$, $(U_v)^{R_v}$ and $(U_x)^{R_x}$  for which we use shared boolean functions/common randomness to decode.
\end{itemize}

We assumed that $NR_{11} > |I_1^v \cap I_3^v|$. Suppose if that does not hold, then we do not have to perform chaining at the second level. The private message bits corresponding to the rate $R_{11}$ will fit into $I_1^v \cap I_3^v$ and hence can be recovered by receiver-$3$ and receiver-$1$. Allocation of the private message bits in $(U_x)^{1:N}$ corresponding to the rate $R_{12}$ will still be the same as in construction for the previously assumed condition.  Fig. \ref{fig5}  shows the allocation of private and public message bits in $(U_w)^{1:N}$ and $(U_v)^{1:N}$ in the chaining procedure for $k=3$ when  $NR_{11} \leq |I_1^v \cap I_3^v|$. The other details of the construction can easily be extended from the construction under the assumption $NR_{11} > |I_1^v \cap I_3^v|$.  Fig. \ref{fig5} shows a case where the public message bits in $I_3^v$ 
 fit into $I_3^v \cap F_1^v$.  Notice that the same chaining procedure still applies, as shown in Fig. \ref{fig5} even when these public message bits overflow into  $I_3^v \cap I_1^v$.
 
 \begin{figure}[t]
\centering
\includegraphics[scale=0.45]{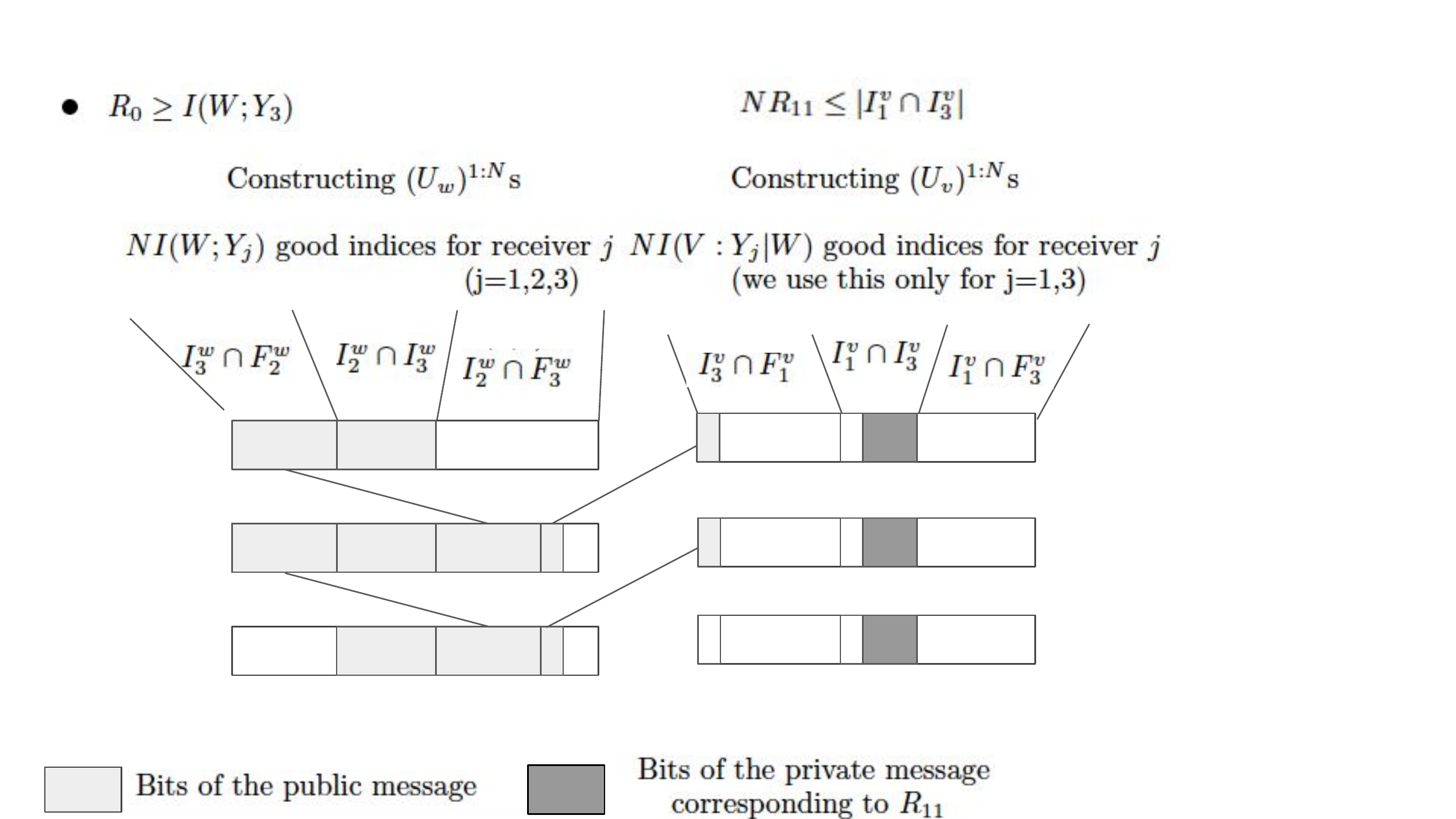}
\caption{Private and public message bits allocation in $(U_w)^{1:N}$ and  $(U_v)^{1:N}$  vectors when $k=3$ }
\label{fig5}
\end{figure}

Now we look at the other case where $R_0 < I(W;Y_3)$. Assume  $NR_0 > |I_2^w \cap I_3^w|$ where there will be non-trivial chaining construction. In this case, note that $NR_0$ public message bits totally fit  into $|I_3^w|$. We  perform chaining within the layer $(U_w)^{1:N}$ itself and resolve the alignment of bit-channel sets $I_2^w$ and  $I_3^w$ so that these public message bits can be reliably decoded at all the receivers. Since we do not require to fill the public message bits in $I_3^v$, receiver-$3$ can ignore decoding the $(U_v)^{1:N}$ vectors. Hence there will be no need of chaining at the second level that aligns bit-channel sets in $I_1^v$ and $I_3^v$ for private message bits corresponding to the rate $R_{11}$. It is just enough to provide private message bits in $I_1^v$. The other details of the code construction can easily be extended from earlier case. Fig. \ref{fig1} shows the case under the assumption $NR_0 > |I_2^w \cap I_3^w|$ when $k=3$.

Suppose if $NR_0 \leq |I_2^w \cap I_3^w|$, we can fill $NR_0$  public message bits in $I_2^w \cap I_3^w$ so that they can be recovered at all the receivers. We can the fill $NR_{11}$ and $NR_{12}$ private message bits in $I_1^v$ of $(U_v)^{1:N}$ and $I_1^x$ of $(U_x)^{1:N}$  so that they can be reliably decoded at receiver-$1$. Hence chaining is not needed when   $NR_0 \leq |I_2^w \cap I_3^w|$.
\begin{figure}[t]
\centering
\includegraphics[scale=0.45]{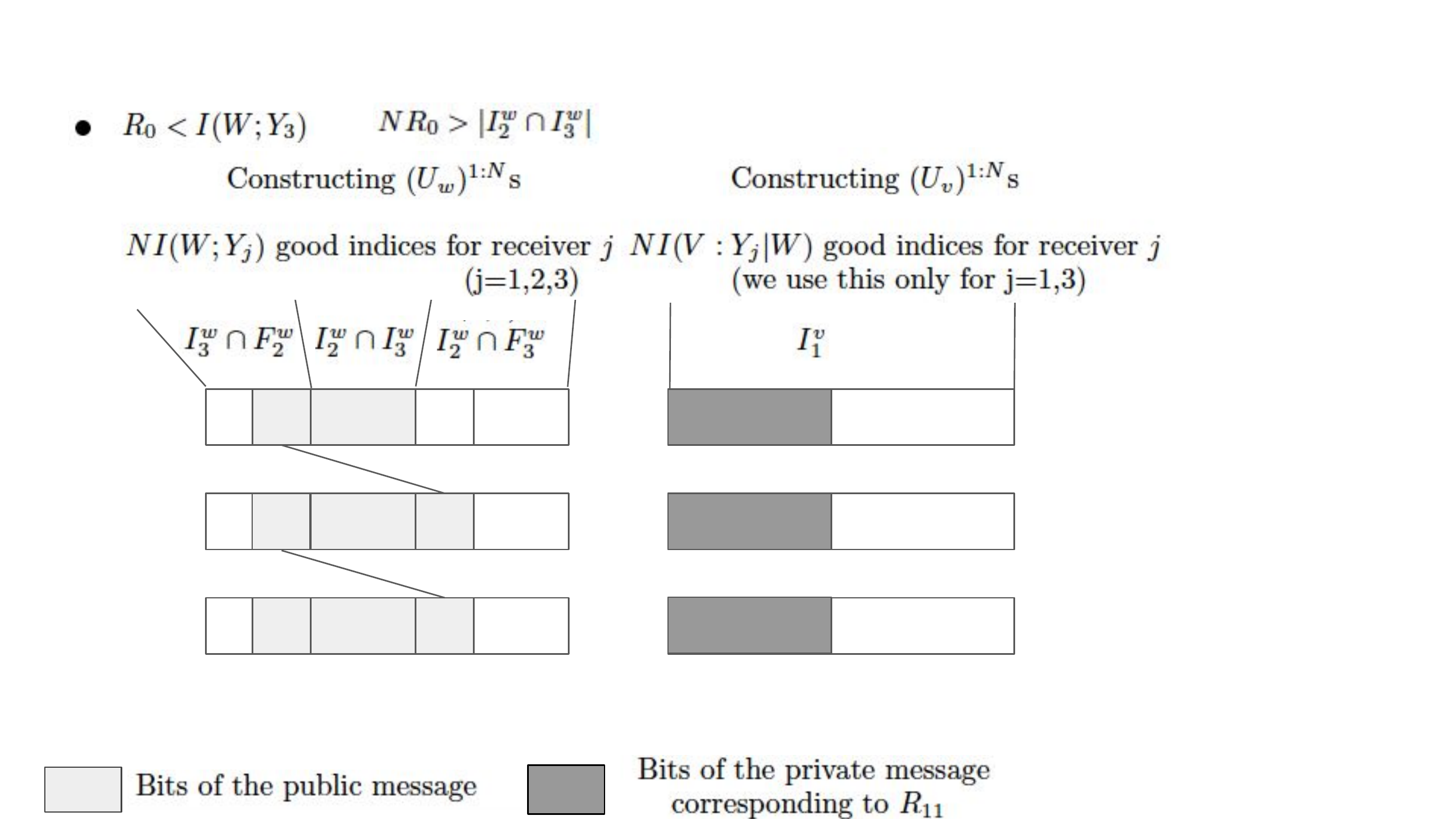}
\caption{Private and public message bits allocation in $(U_w)^{1:N}$ and  $(U_v)^{1:N}$  vectors when $k=3$ }
\label{fig1}
\end{figure}

\subsection{Probability of error analysis}

Let $\mathbbm{C}$ denotes the random variable which contains randomly chosen frozen bits in the code construction. The random variable $\mathbbm{C}$ also contains randomly chosen boolean functions for not-completely polarized bit-channels in case we do not employ common randomness in the code construction, of all the blocks as its components. We give the analysis for the code construction that uses common randomness for not-completely polarized bit-channels. The spirit of the analysis will remain the same for the case where we use randomly chosen boolean functions for encoding not completely polarized bit-channels as in \cite{honda}. The analysis of probability of error that we provide is done in three steps. First step is  deriving the average distribution of each block which is close to the distribution induced when $(W^{1:N}, V^{1:N}, X^{1:N})$ is i.i.d. according to $p(w)p(v|w)p(x|v)$ in total variation distance through Lemma \ref{lemma:3}.  Secondly, we write error event at each receiver as a union of error events we define for each of these blocks. Notice that blocks involved in the chaining are statistically dependent due to the chaining construction we did. We use linearity of expectation and union bound to get an upper bound on average probability of error at a receiver, which is sum of average probability of errors of each of these blocks at that receiver.  Finally,  we use the fact that the total variation distance between average distribution of the block in the code construction and distribution when $(W^{1:N}, V^{1:N}, X^{1:N})$ is i.i.d. according to $p(w)p(v|w)p(x|v)$ are close and polarization results to get bound on the average probability of each block at that receiver.  Theorem \ref{theorem:1} provides a detailed analysis of the probability of decoding error for the chaining construction. We now give Lemma \ref{lemma:3} and Lemma \ref{lemma:4} used in Theorem \ref{theorem:1}.
\begin{lemma}\label{lemma:3}
Let $Q_{(U_w)^{1:N} (U_v)^{1:N} (U_x)^{1:N}}$ be the measure on $(U_w)^{1:N}(U_v)^{1:N} (U_x)^{1:N}$ as follows:
\\
 $Q_{(U_w)^{1:N} (U_v)^{1:N}(U_x)^{1:N}}(u_w^{1:N} u_v^{1:N} u_x^{1:N})      \\ \text{\hspace{0.2cm}}=
\big(2^{-|\mathcal{H}_W|} \Pi_{i \in \mathcal{L}_W} \delta_i^w((u_w)_i|(u_w)^{1:i-1}) \\ \text{\hspace{2.7cm}}   \Pi_{i \in R^w}P_{(U_w)_i|(U_w)^{1:{i-1}}} ((u_w)_i|(u_w)^{1:i-1})\big)\cdot
\text{\hspace{0.7cm}}
\big(2^{-|\mathcal{H}_{V|W}|}  \Pi_{i \in \mathcal{L}_{V|W}} \delta_i^v((u_v)_i|w^{1:N} (u_v)^{1:i-1} ) \\ \text{\hspace{1.5cm}}\Pi_{i \in R^v} P_{(U_v)_i|W^{1:N} (U_v)^{1:{i-1}} }  ( (u_v)_i|w^{1:N} (u_v)^{1:i-1} )\big)\cdot
\\\text{\hspace{0.7cm}}
\big(2^{-|\mathcal{H}_{X|V}|} \Pi_{i \in \mathcal{L}_{X|V}} \delta_i^x((u_x)_i|v^{1:N} (u_x)^{1:i-1} ) \\ \text{\hspace{1.5cm}}\Pi_{i \in R^x} P_{(U_x)_i|V^{1:N} (U_x)^{1:{i-1} }} ((u_x)_i|v^{1:N} (u_x)^{1:i-1}  ) \big).$\\
Let $P_{(U_w)^{1:N}(U_v)^{1:N} (U_x)^{1:N}}$ be the measure induced when $(W^{1:N}, V^{1:N}, X^{1:N})$ are i.i.d.  according to $p(w)p(v|w)p(x|v)$. The total variation distance, $ || P_{(U_w)^{1:N}(U_v)^{1:N} (U_x)^{1:N}} - Q_{(U_w)^{1:N}(U_v)^{1:N} (U_x)^{1:N}}
|| = O(2^{-N^{\beta'}})$, where $\beta' < \beta < 0.5$, $w^N = (u_w)^{1:N}G_N$, $v^N = (u_v)^{1:N}G_N$ and  $x^N = (u_x)^{1:N}G_N$.
\end{lemma}
\textbf{Proof:}\\
We use short hand notation  $Q ((u_w)^{1:N} (u_v)^{1:N} (u_x)^{1:N})$ for $Q_{(U_w)^{1:N} (U_v)^{1:N}(U_x)^{1:N}}((u_w)^{1:N} (u_v)^{1:N} (u_x)^{1:N})$. Similarly, we use short hand notation  $P ((u_w)^{1:N} (u_v)^{1:N} (u_x)^{1:N})$ for $P_{(U_w)^{1:N} (U_v)^{1:N}(U_x)^{1:N}}((u_w)^{1:N} (u_v)^{1:N} (u_x)^{1:N})$. The proof is inspired from Lemma 1 in~\cite{honda}. From equation (56) in  ~\cite{honda}, we have the following identity:
\begin{equation}\label{eq:13}
B_1^n - A_1^n 
= \sum_{i=1}^n (B_i - A_i) A_1^{i-1} B_{i+1}^n
\end{equation}
where $A_j^k$ and $B_j^k$ denotes the product $\prod_{i=j}^k A_i$ and $\prod_{i=j}^k B_i$ respectively. We are going to apply this for $n=3N$ length vector, which is $(U_w)^{1:N} (U_v)^{1:N}(U_x)^{1:N}$.
\begin{align*}
&2 ||Q_{(U_w)^{1:N} (U_v)^{1:N}(U_x)^{1:N}} - P_{(U_w)^{1:N} (U_v)^{1:N}(U_x)^{1:N}}||\\
&= \sum_{(u_w)^{1:N} (u_v)^{1:N} (u_x)^{1:N}} \\& \text{\hspace{0.5cm }}  |\big( \sum_{i=1}^N   P((u_w)_i| (u_w)^{1:i-1}) -  Q((u_w)_i|  (u_w)^{1:i-1}) ) \\&\text{\hspace{1.7cm}} \Pi_{m=1}^{i-1} P((u_w)_m| (u_w)^{1:m-1})\\&\text{\hspace{1.9cm}} \Pi_{m=i+1}^N Q((u_w)_m| (u_w)^{1:m-1}) 
\\&\text{\hspace{2.0cm}} \Pi_{p=1}^N Q((u_v)_p| w^{1:N} (u_v)^{1:p-1}) 
\\&\text{\hspace{2.1cm}} \Pi_{q=1}^N Q((u_x)_q| w^{1:N} v^{1:N} (u_x)^{1:q-1}) 
\\& \text{\hspace{0.1cm }} +  \sum_{k=1}^N   (P((u_v)_k| w^{1:N} (u_v)^{1:k-1}) -  Q((u_v)_k| w^{1:N} (u_v)^{1:k-1}))  \\&\text{\hspace{1.7cm}} \Pi_{m=1}^N P((u_w)_m| (u_w)^{1:m-1})\\&\text{\hspace{1.9cm}} \Pi_{p=1}^{k-1} P((u_v)_p| w^{1:N}(u_v)^{1:p-1}) 
\\&\text{\hspace{2.0cm}} \Pi_{p=k+1}^N Q((u_v)_p| w^{1:N} (u_v)^{1:p-1}) 
\\&\text{\hspace{2.1cm}} \Pi_{q=1}^N Q((u_x)_q| w^{1:N} v^{1:N} (u_x)^{1:q-1}) 
\\& \text{\hspace{0.1cm }} +  \sum_{l=1}^N   (P((u_x)_l|w^{1:N} v^{1:N} (u_x)^{1:l-1}) - \\& \text{\hspace{4cm }}  Q((u_x)_k| w^{1:N} v^{1:N} (u_x)^{1:k-1}))  \\&\text{\hspace{1.7cm}} \Pi_{m=1}^N P((u_w)_m| (u_w)^{1:m-1})\\&\text{\hspace{1.9cm}} \Pi_{p=1}^{N} P((u_v)_p| w^{1:N}(u_v)^{1:p-1}) 
\\&\text{\hspace{2.0cm}} \Pi_{q=1}^{l-1} P((u_x)_q|  w^{1:N} v^{1:N} (u_x)^{1:q-1})) 
\\&\text{\hspace{2.1cm}} \Pi_{q=1}^N Q((u_x)_q| w^{1:N} v^{1:N} (u_x)^{1:q-1}) \big)|
\end{align*}
This implies that 
\begin{align*}
&2 ||Q_{(U_w)^{1:N} (U_v)^{1:N}(U_x)^{1:N}} - P_{(U_w)^{1:N} (U_v)^{1:N}(U_x)^{1:N}}||\\
&\leq \sum_{(u_w)^{1:N} (u_v)^{1:N} (u_x)^{1:N}} \\& \text{\hspace{0.5cm }}  \big( \sum_{i=1}^N  | P((u_w)_i| (u_w)^{1:i-1}) -  Q((u_w)_i|  (u_w)^{1:i-1}) )| \\&\text{\hspace{1.2cm}}  P((u_w)^{1:i-1}) Q((u_w)^{i+1:N} (u_v)^{1:N} (u_x)^{1:N}| (u_w)^{1:i}) 
\\& +  \sum_{k=1}^N   |(P((u_v)_k| w^{1:N} (u_v)^{1:k-1}) -  \\& \text{\hspace{4cm }}  Q((u_v)_k| w^{1:N} (u_v)^{1:k-1}))|    \\&\text{\hspace{1.2cm}}  P((u_w)^{1:N} (u_v)^{1:k-1})\\&\text{\hspace{1.4cm}} 
Q((u_v)^{k+1:N} (u_x)^{1:N}| (w)^{1:N}(u_v)^{1:k})
\\& \text{\hspace{0.1cm }} +  \sum_{l=1}^N   |(P((u_x)_l| v^{1:N} (u_x)^{1:l-1}) - \\& \text{\hspace{4cm }}  Q((u_x)_k|  v^{1:N} (u_x)^{1:k-1})) | \\&\text{\hspace{1.2cm}}  P((u_w)^{1:N} (u_v)^{1:N} (u_x)^{1:l-1})\\&\text{\hspace{1.4cm}} 
Q( (u_x)^{l+1:N}| (w)^{1:N}(v)^{1:N}) (u_x)^{l} \big)
\end{align*}
This implies that 
\begin{align*}
&2 ||Q_{(U_w)^{1:N} (U_v)^{1:N}(U_x)^{1:N}} - P_{(U_w)^{1:N} (U_v)^{1:N}(U_x)^{1:N}}||\\
&\leq \big ( \sum_{(u_w)^{1:N} (u_v)^{1:N} (u_x)^{1:N}} \\& \text{\hspace{0.5cm }}   \sum_{i=1}^N  | P((u_w)_i| (u_w)^{1:i-1}) -  Q((u_w)_i|  (u_w)^{1:i-1}) )| \\&\text{\hspace{1.2cm}}  P((u_w)^{1:i-1}) Q((u_w)^{i+1:N} (u_v)^{1:N} (u_x)^{1:N}| (u_w)^{1:i}) \big) 
\\
&+ \big ( \sum_{(u_w)^{1:N} (u_v)^{1:N} (u_x)^{1:N}} \\&
   \sum_{k=1}^N   |(P((u_v)_k| w^{1:N} (u_v)^{1:k-1}) -  Q((u_v)_k| w^{1:N} (u_v)^{1:k-1}))| 
\\&\text{\hspace{1.2cm}}  P((u_w)^{1:N} (u_v)^{1:k-1})\\&\text{\hspace{1.4cm}} 
Q((u_v)^{k+1:N} (u_x)^{1:N}| (w)^{1:N}(u_v)^{1:k-1})\big)
\\
&+ \big ( \sum_{(u_w)^{1:N} (u_v)^{1:N} (u_x)^{1:N}} \\&
 \text{\hspace{0.1cm }}   \sum_{l=1}^N   |(P((u_x)_l| v^{1:N} (u_x)^{1:l-1}) - \\& \text{\hspace{4cm }}  Q((u_x)_k|  v^{1:N} (u_x)^{1:k-1})) | \\&\text{\hspace{1.2cm}}  P((u_w)^{1:N} (u_v)^{1:N} (u_x)^{1:l-1})\\&\text{\hspace{1.4cm}} 
Q( (u_x)^{l+1:N}|  (w)^{1:N}(v)^{1:N} (u_x)^{l} \big)
\end{align*}
This implies that 

\begin{align*}
&2 ||Q_{(U_w)^{1:N} (U_v)^{1:N}(U_x)^{1:N}} - P_{(U_w)^{1:N} (U_v)^{1:N}(U_x)^{1:N}}||\\
&\leq \big (   \sum_{i=1}^N  \sum_{(u_w)^{1:N} (u_v)^{1:N} (u_x)^{1:N}} \\& \text{\hspace{0.5cm }} | P((u_w)_i| (u_w)^{1:i-1}) -  Q((u_w)_i|  (u_w)^{1:i-1}) )| \\&\text{\hspace{1.2cm}}  P((u_w)^{1:i-1}) Q((u_w)^{i+1:N} (u_v)^{1:N} (u_x)^{1:N}| (u_w)^{1:i}) \big) 
\\
&+ 
   \big ( \sum_{k=1}^N   \sum_{(u_w)^{1:N} (u_v)^{1:N} (u_x)^{1:N}} \\& \text{\hspace{1cm}} |(P((u_v)_k| w^{1:N} (u_v)^{1:k-1}) -  Q((u_v)_k| w^{1:N} (u_v)^{1:k-1}))| 
\\&\text{\hspace{1.2cm}}  P((u_w)^{1:N} (u_v)^{1:k-1})\\&\text{\hspace{1.4cm}} 
Q((u_v)^{k+1:N} (u_x)^{1:N}| (w)^{1:N}(u_v)^{1:k-1})\big)
\\
&+ \big ( 
 \text{\hspace{0.1cm }}   \sum_{l=1}^N \sum_{(u_w)^{1:N} (u_v)^{1:N} (u_x)^{1:N}} \\& \hspace{1cm} |(P((u_x)_l| v^{1:N} (u_x)^{1:l-1}) - \\& \text{\hspace{4cm }}  Q((u_x)_k|  v^{1:N} (u_x)^{1:k-1})) | \\&\text{\hspace{1.2cm}}  P((u_w)^{1:N} (u_v)^{1:N} (u_x)^{1:l-1})\\&\text{\hspace{1.4cm}} 
Q( (u_x)^{l+1:N}| (w)^{1:N}(v)^{1:N} (u_x)^{l} \big)
\end{align*}
This implies that
\begin{align}\label{eq:14}
\begin{split}
&2 ||Q_{(U_w)^{1:N} (U_v)^{1:N}(U_x)^{1:N}} - P_{(U_w)^{1:N} (U_v)^{1:N}(U_x)^{1:N}}||\\
&\leq \big (   \sum_{i=1}^N  \sum_{(u_w)^{1:i} } P((u_w)^{1:i-1}) \\& \text{\hspace{1.5cm }} | P((u_w)_i| (u_w)^{1:i-1}) -  Q((u_w)_i|  (u_w)^{1:i-1}) )|    \big) 
\\
&+ 
  \big ( \sum_{k=1}^N   \sum_{(u_w)^{1:N} (u_v)^{1:k} } P((u_w)^{1:N} (u_v)^{1:k-1}) \\& \text{\hspace{1cm}} |(P((u_v)_k| w^{1:N} (u_v)^{1:k-1}) -  \\& \text{\hspace{4cm }}  Q((u_v)_k| w^{1:N} (u_v)^{1:k-1}))| 
  \big)
\\
&+ \big ( 
 \text{\hspace{0.1cm }}   \sum_{l=1}^N \sum_{(u_w)^{1:N} (u_v)^{1:N} (u_x)^{1:l}} P((u_w)^{1:N} (u_v)^{1:N} (u_x)^{1:l-1}) \\& \hspace{1cm} |(P((u_x)_l| v^{1:N} (u_x)^{1:l-1}) - \\& \text{\hspace{4cm }}  Q((u_x)_k|  v^{1:N} (u_x)^{1:k-1})) |  
 \big)
 \end{split}
\end{align}
Now we consider the individual sum terms in the above bound. Let us first bound the term, \\$   \sum_{i=1}^N  \sum_{(u_w)^{1:i} } P((u_w)^{1:i-1}) \\ \text{\hspace{1.5cm }} | P((u_w)_i| (u_w)^{1:i-1}) -  Q((u_w)_i|  (u_w)^{1:i-1}) )|     $. \\

If $i \in \mathcal{H}_W $, then 
\begin{align*}
&\sum_{(u_w)^{1:i} } P((u_w)^{1:i-1})  | P((u_w)_i| (u_w)^{1:i-1})  \\&
\text{\hspace{2cm}} -Q((u_w)_i|  (u_w)^{1:i-1}) )|\\ &= \sum_{(u_w)^{1:i-1}} 2 P((u_w)^{1:i-1})  || P_{(U_w)_i| (U_w)^{1:i-1} = (u_w)^{1:i-1}} - \\&
\text{\hspace{1.5cm}} Q_{(U_w)_i| (U_w)^{1:i-1} = (u_w)^{1:i-1}} ||\\
& 
\stackrel{(a)}\leq \sum_{(u_w)^{1:i-1}}P((u_w)^{1:i-1}) \sqrt{(2\ln2)} \\& \text{\hspace{0.4cm}} \big ( D(P_{(U_w)_i|(U_w)^{1:i-1} =(u_w)^{1:i-1}} ||\\& \text{\hspace{2.5cm}} Q_{(U_w)_i|(U_w)^{1:i-1} =(u_w)^{1:i-1}})\big)^{0.5}\\
&\stackrel{(b)}\leq  \sqrt{(2\ln 2)}  \big ( \sum_{(u_w)^{1:i-1}} P((u_w)^{1:i-1}) \\ & \text{\hspace{0.4cm}} D(P_{(U_w)_i|(U_w)^{1:i-1} =(u_w)^{1:i-1}} ||\\& \text{\hspace{2.5cm}} Q_{(U_w)_i|(U_w)^{1:i-1} =(u_w)^{1:i-1}})\big)^{0.5}\\
&\stackrel{(c)}\leq  \sqrt{(2\ln 2)  (1 - H((U_w)_i | (U_w)^{1:i-1}))}\\
& \stackrel{(d)}\leq \sqrt{(2\ln 2)  (1 - (Z((U_w)_i | (U_w)^{1:i-1}))^2)}\\
& \stackrel{(e)}\leq  \sqrt{(4\ln 2)  (2^{-n^\beta})}\\
&=O(2^{-n^{\beta'}})
\end{align*}
where $\beta' < \beta$.
\\ (a) follows by pinsker inequality, (b) follows by jensen's inequality, (c) follows due to the fact that  $Q((u_w)_i|(u_w)^{1:i-1}) = 0.5$ and by the formula of conditional entropy, (d) follows from equation (\ref{eq:15}), and  (e) follows  from polarization results mentioned in Section \ref{np}.\\
If $i \in \mathcal{L}_W$, 
let \[ p_{(u_w)^{1:i-1}} = \max \{P(0|(u_w)^{1:i-1}), P(1|(u_w)^{1:i-1}) \} \]
Then, 
\begin{align*}
&\sum_{(u_w)^{1:i} } P((u_w)^{1:i-1})  | P((u_w)_i| (u_w)^{1:i-1})  \\&
\text{\hspace{2cm}} -Q((u_w)_i|  (u_w)^{1:i-1}) )|\\ &= \sum_{(u_w)^{1:i-1}} 2 P((u_w)^{1:i-1})  || P_{(U_w)_i| (U_w)^{1:i-1} = (u_w)^{1:i-1}} - \\&
\text{\hspace{1.5cm}} Q_{(U_w)_i| (U_w)^{1:i-1} = (u_w)^{1:i-1}} ||\\
& 
\stackrel{(a)}\leq \sum_{(u_w)^{1:i-1}}P((u_w)^{1:i-1}) \sqrt{(2\ln2)} \\& \text{\hspace{0.5cm}} \big(D(Q_{(U_w)_i|(U_w)^{1:i-1} =(u_w)^{1:i-1}} || \\& \text{\hspace{3.5cm}} P_{(U_w)_i|(U_w)^{1:i-1} =(u_w)^{1:i-1}})\big)^{0.5}\\
&\stackrel{(b)}\leq  \sqrt{(2\ln 2)}  \big ( \sum_{(u_w)^{1:i-1}} P((u_w)^{1:i-1}) \\ & \text{\hspace{0.4cm}} D(Q_{(U_w)_i|(U_w)^{1:i-1} =(u_w)^{1:i-1}} || \\& \text{\hspace{3.5cm}} P_{(U_w)_i|(U_w)^{1:i-1} =(u_w)^{1:i-1}})\big)^{0.5}\\
&\stackrel{(c)}\leq  \sqrt{(2\ln 2) \sum_{(u_w)^{1:i-1}} P((u_w)^{1:i-1}) (-\log(p_{(u_w)^{1:i-1}}))}\\
&\stackrel{(d)}\leq  \big((2\ln 2) \sum_{(u_w)^{1:i-1}} P((u_w)^{1:i-1}) \\& \text{\hspace{2.5cm}}(H((U_w)_i| (U_w)^{1:i-1} = (u_w)^{1:i-1}))\big)^{0.5}\\
&=   \sqrt{(2\ln 2)  (H((U_w)_i| (U_w)^{1:i-1} ))}\\
&\stackrel{(e)} 
\leq \sqrt{(2\ln 2)  (Z((U_w)_i| (U_w)^{1:i-1} ))}\\
& \stackrel{(f)}\leq \sqrt{(2\ln 2) 2^{-n^\beta}} = O(2^{-n^{\beta'}})
\end{align*}
(a) follows by pinsker inequality, (b) follows by  jensen's inequality for concave functions. (c) follows from $Q((u_w)_i|(u_w)^{1:i-1}) = 1$ when $(u_w)_i = \text{argmax}_{x \in \{0,1\}}  \{P(x|(u_w)^{1:i-1})\}$. (d) is true since $\log(\frac{p_{(u_w)^{1:i-1}}}{1- p_{(u_w)^{1:i-1}}}) > 0$. (e) follows from equation (\ref{eq:16}), (f) follows from polarization results mentioned in Section \ref{np}.

Hence 
\begin{align}
\begin{split}\label{eq:17}
&\sum_{i=1}^N  \sum_{(u_w)^{1:i} } P((u_w)^{1:i-1}) \\& \text{\hspace{0.2cm }} | P((u_w)_i| (u_w)^{1:i-1}) -  Q((u_w)_i|  (u_w)^{1:i-1}) )|   = O(2^{-N^{\beta'}}) .
\end{split}
\end{align}
 Let us first bound the term, \\$   \sum_{k=1}^N  \sum_{(u_w)^{1:N} (u_v)^{1:k} } P((u_v)^{1:k-1})  |P((u_v)_k| w^{1:N} (u_v)^{1:k-1}) \\ \text{\hspace{3.5cm }} -Q((u_v)_k|w^{1:N}  (u_w)^{1:k-1}) )|     $. \\

If $i \in \mathcal{H}_{V|W} $, then 
\begin{align*}
&\sum_{(u_w)^{1:N}(u_v)^{1:k} } P((u_w)^{1:N} (u_v)^{1:k-1})  | P((u_v)_k| w^{1:N}(u_v)^{1:k-1})  \\&
\text{\hspace{2.5cm}} -Q((u_v)_k| w^{1:N}(u_v)^{1:k-1}) )|\\ &= \sum_{w^{1:N}(u_v)^{1:k-1}} 2 P(w^{1:N} (u_v)^{1:i-1}) \\&
\text{\hspace{1.5cm}} || P_{(U_v)_k| W^{1:N}(U_v)^{1:k-1} = w^{1:N}(u_v)^{1:k-1}} - \\&
\text{\hspace{2.0cm}} Q_{(U_v)_k| W^{1:N}(U_v)^{1:k-1} = w^{1:N}(u_v)^{1:k-1}} ||\\
& 
\stackrel{(a)}\leq \sum_{w^{1:N}(u_v)^{1:k-1}} 2 P(w^{1:N} (u_v)^{1:i-1})  \sqrt{(2\ln2)} \\& \text{\hspace{0.5cm}} \big({ D(P_{(U_v)_k| W^{1:N}(U_v)^{1:k-1} = w^{1:N}(u_v)^{1:k-1}} } ||\\& \text{\hspace{2.5cm}} Q_{(U_v)_k| W^{1:N}(U_v)^{1:k-1} = w^{1:N}(u_v)^{1:k-1}} \big)^{0.5}\\
&\stackrel{(b)}\leq  \sqrt{(2\ln 2)}  \big ( \sum_{w^{1:N}(u_v)^{1:k-1}} 2 P(w^{1:N} (u_v)^{1:i-1}) \\ & \text{\hspace{0.2cm}} \big({ D(P_{(U_v)_k| W^{1:N}(U_v)^{1:k-1} = w^{1:N}(u_v)^{1:k-1}} } ||\\& \text{\hspace{2.5cm}} Q_{(U_v)_k| W^{1:N}(U_v)^{1:k-1} = w^{1:N}(u_v)^{1:k-1}} \big)^{0.5}\\
&\stackrel{(c)}\leq  \sqrt{(2\ln 2)  (1 - H((U_v)_k | W^{1:N}(U_v)^{1:k-1}))}\\
& \stackrel{(d)}\leq \sqrt{(2\ln 2)  (1 - (Z((U_v)_k | W^{1:N}(U_v)^{1:k-1}))^2)}\\
& \stackrel{(e)}\leq  \sqrt{(4\ln 2)  (2^{-n^\beta})}\\
&=O(2^{-n^{\beta'}})
\end{align*}
where $\beta' < \beta$.
\\ (a) follows by pinsker inequality, (b) follows by jensen's inequality, (c) follows due to the fact that  $Q((u_v)_k|w^{1:N}(u_v)^{1:k-1}) = 0.5$ and by the formula of conditional entropy, (d) follows from equation (\ref{eq:15}) and  (e) follows  from polarization results mentioned in Section \ref{np}.\\
Let $ p_{w^{1:N}(u_v)^{1:k-1}}=\\ \text{\hspace{1.5cm}}  \max \{P(0|w^{1:N} (u_v)^{1:k-1}), P(1|w^{1:N}(u_v)^{1:k-1}) \}$.\\
If $i \in \mathcal{L}_{V|W}$,
then, 
\begin{align*}
&\sum_{(u_w)^{1:N}(u_v)^{1:k} } P((u_w)^{1:N} (u_v)^{1:k-1})  | P((u_v)_k| w^{1:N}(u_v)^{1:k-1})  \\&
\text{\hspace{2.5cm}} -Q((u_v)_k| w^{1:N}(u_v)^{1:k-1}) )|\\ &= \sum_{w^{1:N}(u_v)^{1:k-1}} 2 P(w^{1:N} (u_v)^{1:i-1}) \\&
\text{\hspace{1.5cm}} || P_{(U_v)_k| W^{1:N}(U_v)^{1:k-1} = w^{1:N}(u_v)^{1:k-1}} - \\&
\text{\hspace{2.0cm}} Q_{(U_v)_k| W^{1:N}(U_v)^{1:k-1} = w^{1:N}(u_v)^{1:k-1}} ||\\
& 
\stackrel{(a)}\leq \sum_{w^{1:N}(u_v)^{1:k-1}} 2 P(w^{1:N} (u_v)^{1:i-1})  \sqrt{(2\ln2)} \\& \text{\hspace{0.5cm}} \big({ D(Q_{(U_v)_k| W^{1:N}(U_v)^{1:k-1} = w^{1:N}(u_v)^{1:k-1}} } ||\\& \text{\hspace{2.5cm}} P_{(U_v)_k| W^{1:N}(U_v)^{1:k-1} = w^{1:N}(u_v)^{1:k-1}} \big)^{0.5}\\
&\stackrel{(b)}\leq  \sqrt{(2\ln 2)}  \big ( \sum_{w^{1:N}(u_v)^{1:k-1}} 2 P(w^{1:N} (u_v)^{1:i-1}) \\ & \text{\hspace{0.2cm}} \big({ D(Q_{(U_v)_k| W^{1:N}(U_v)^{1:k-1} = w^{1:N}(u_v)^{1:k-1}} } ||\\& \text{\hspace{2.5cm}} P_{(U_v)_k| W^{1:N}(U_v)^{1:k-1} = w^{1:N}(u_v)^{1:k-1}} \big)^{0.5}\\
&\stackrel{(c)}\leq  \sqrt{(2\ln 2)}\\& \text{\hspace{0.5cm}}\big( \sum_{(u_w)^{1:N}(u_v)^{1:k} } P((u_w)^{1:N} (u_v)^{1:k-1})\\& \text{\hspace{4.5cm}} (-\log(p_{w^{1:N}(u_v)^{1:k-1}}))\big)^{0.5}\\
&\stackrel{(d)}\leq  \sqrt{(2\ln 2)}\\& \text{\hspace{0.5cm}}\big( \sum_{(u_w)^{1:N}(u_v)^{1:k} } P((u_w)^{1:N} (u_v)^{1:k-1})\\& \text{\hspace{1.5cm}}(H((U_v)_i| W^{1:N}(U_v)^{1:i-1} = w^{1:N}(u_v)^{1:i-1}))\big)^{0.5}\\
&=   \sqrt{(2\ln 2)  (H((U_v)_i|W^{1:N} (U_v)^{1:i-1} ))}\\
&\stackrel{(e)} 
\leq \sqrt{(2\ln 2)  (Z((U_v)_i| W^{1:N}(U_v)^{1:i-1} ))}\\
& \stackrel{(f)}\leq \sqrt{(2\ln 2) 2^{-n^\beta}} = O(2^{-n^{\beta'}})
\end{align*}
(a) follows by pinsker inequality, (b) follows by  jensen's inequality for concave functions. (c) follows from $Q((u_w)_i|(u_w)^{1:i-1}) = 1$ when $(u_v)_k = \text{argmax}_{x\in \{0,1\}}  P(x|w^{1:N}(u_v)^{1:k-1}) $. (d) is true since $\log(\frac{p_{w^{1:N}(u_v)^{1:k-1}}}{1- p_{w^{1:N}(u_v)^{1:k-1}}}) > 0$. (e) follows from equation (\ref{eq:16}), (f) follows from equation (\ref{eq:16}), (f) follows from polarization results mentioned in Section \ref{np}.
Hence
\begin{align}\label{eq:18}
\begin{split}
  & \sum_{k=1}^N  \sum_{(u_w)^{1:N} (u_v)^{1:k} } P((u_v)^{1:k-1})  |P((u_v)_k| w^{1:N} (u_v)^{1:k-1}) \\& \text{\hspace{2.5cm }} -Q((u_v)_k|w^{1:N}  (u_w)^{1:k-1}) )| = O(2^{-N^{\beta'}})  
  \end{split}
\end{align}
By using the same approach as we used to derive equation $(\ref{eq:18})$, we will also get
\begin{align}\label{eq:19}
\begin{split}
  & \sum_{l=1}^N  \sum_{(u_w)^{1:N} (u_v)^{1:N} (u_x)^{1:l}} P((u_x)^{1:l-1})  |P((u_x)_l| v^{1:N} (u_x)^{1:l-1}) \\& \text{\hspace{2.5cm }} -Q((u_x)_l|v^{1:N}  (u_x)^{1:l-1}) )| = O(2^{-N^{\beta'}})   
 \end{split}
\end{align}
From equations $(\ref{eq:14})$, $(\ref{eq:17})$, $(\ref{eq:18})$ and $(\ref{eq:19})$, we get $ ||P_{(U_w)^{1:N}(U_v)^{1:N} (U_x)^{1:N}} - Q_{(U_w)^{1:N}(U_v)^{1:N} (U_x)^{1:N}}
|| = O(2^{-N^{\beta'}}).$ Hence proof of the lemma.\qed

\begin{lemma}\label{lemma:4}
Let the $(X,Y)$ random variable pair have two measures defined as $Q_{X,Y}(x,y) = Q_X(x)p(y|x)$ and $P_{X,Y}(x,y) = P_X(x)p(y|x)$, respectively. So the conditional distributions $Q_{Y|X}(y|x)$ and $P_{Y|X}(y|x)$ are both equal to $p(y|x)$. The total variation between the joint distributions $||Q_{X,Y}
- P_{X,Y}||$ becomes $||Q_X - P_X||$.
\end{lemma}
\textbf{Proof:} 
\begin{align*}
&||Q_{X,Y} - P_{X,Y}|| \\&= \sum_{(x,y): P_{X,Y}(x,y) > Q_{X,Y}(x,y)} P_{X,Y}(x,y) - Q_{X,Y}(x,y)\\
&= \sum_{(x,y): P_X(x)p(y|x)  > Q_X(x)p(y|x)} P_X(x)p(y|x) - Q_X(x)p(y|x)\\
&= \sum_{(x,y): P_X(x) > Q_X(x)} (P_X(x) - Q_X(x)) p(y|x)\\
&= \sum_{x: P_X(x) > Q_X(x)}\sum_{y}(P_X(x) - Q_X(x)) p(y|x)\\
&= \sum_{x: P_X(x) > Q_X(x)}(P_X(x) - Q_X(x))\\
&= ||Q_{X} - P_{X}||. \qed
\end{align*}

Now we provide Theorem \ref{theorem:1} that gives a detailed analysis of the probability of decoding error in the chaining construction. 

\begin{theorem}\label{theorem:1}
\text{\hspace{0.1cm}}\\1. For every polar block encoded in the chaining construction, we have\\
 $\mathbb{E}_{\mathbbm{C}}[{\mathbb{P}(U_w^{1:N}= u_w^{1:N}, U_v^{1:N}= u_v^{1:N}, U_x^{1:N}= u_x^{1:N} |  \mathbbm{C})}] . \\\text{\hspace{0.2cm}}=
\big(2^{-|\mathcal{H}_W|} \Pi_{i \in \mathcal{L}_W} \delta_i^w((u_w)_i|(u_w)^{1:i-1}) \\ \text{\hspace{2.7cm}}   \Pi_{i \in R^w}P_{(U_w)_i|(U_w)^{1:{i-1}}} ((u_w)_i|(u_w)^{1:i-1})\big)\cdot
\text{\hspace{0.7cm}}
\big(2^{-|\mathcal{H}_{V|W}|}  \Pi_{i \in \mathcal{L}_{V|W}} \delta_i^v((u_v)_i|w^{1:N} (u_v)^{1:i-1} ) \\ \text{\hspace{1.5cm}}\Pi_{i \in R^v} P_{(U_v)_i|W^{1:N} (U_v)^{1:{i-1}} }  ( (u_v)_i|w^{1:N} (u_v)^{1:i-1} )\big)\cdot
\\\text{\hspace{0.7cm}}
\big(2^{-|\mathcal{H}_{X|V}|} \Pi_{i \in \mathcal{L}_{X|V}} \delta_i^x((u_x)_i|v^{1:N} (u_x)^{1:i-1} ) \\ \text{\hspace{1.5cm}}\Pi_{i \in R^x} P_{(U_x)_i|V^{1:N} (U_x)^{1:{i-1} }} ((u_x)_i|v^{1:N} (u_x)^{1:i-1}  ) \big).$\\
where $w^{1:N} = (u_w)^{1:N} G_N$, $v^{1:N} = (u_v)^{1:N} G_N$ and $x^{1:N} = (u_x)^{1:N} G_N$. \\
2. Let $P_{e}(\mathbbm{C})$ be the probability of error  for a given code in the above random chaining construction with $k$ blocks. The average probability of error for the random code construction,  $\mathbb{E}_{\mathbbm{C}}[P_{e}(\mathbbm{C})] = O(k2^{-N^{\beta'}})$ for $\beta' < \beta < 0.5$. 
\end{theorem}
\textbf{Proof:}\\
1.\\
Let us consider a polar block in the random chaining construction. We now compute the ensemble average distribution of such a block. We first evaluate ${\mathbb{P}(U_w^{1:N}= u_w^{1:N}|  \mathbbm{C})}$ for that block.\\
Remember that in the code construction, we give the private and public message bits in a portion of $U^{\mathcal{H}_W}$ and we put randomly chosen frozen bits with i.i.d.  uniform distribution in the remaining portion of it. Let $I^w$ be index set where we put private/public message bits in  $U^{\mathcal{H}_W}$ in that block. Let the randomly chosen frozen bit function be  $f_w:\mathcal{H}_W - I_w \rightarrow \{0, 1\}$.  By encoding method, we get, \\
${\mathbb{P}(U_w^{1:N}= u_w^{1:N}|  \mathbbm{C})} \\
\text{\hspace{0.2cm}}= \Pi_{i \in [N]} \mathbb{P}((U_w)_i = (u_w)_i| \mathbbm{C}, (U_w)^{1:i-1}= (u_w)^{1:i-1})\\
\text{\hspace{0.2cm}}= 2^{-|I_w|} \Pi_{i \in \mathcal{H}_W - I} \mathbb{1}\{f_w(i) = w_i\}\\ \text{\hspace{2cm}} \Pi_{i \in \mathcal{L}_W} \delta_i^w((u_w)_i|(u_w)^{1:i-1}) \\ \text{\hspace{2.5cm}}   \Pi_{i \in R^w}P_{(U_w)_i|(U_w)^{1:{i-1}}} ((u_w)_i|(u_w)^{1:i-1})\big)$.\\
By taking expectation on both sides, by independence of frozen bits and by the linearity of expectation, we get the following:\\
$\mathbb{E}_{\mathbbm{C}}[{\mathbb{P}(U_w^{1:N}= u_w^{1:N}|  \mathbbm{C})}]\\ 
\text{\hspace{0.2cm}}= 2^{-|I_w|} \Pi_{i \in \mathcal{H}_W - I} \mathbb{E}_{\mathbbm{C}}[\mathbb{1}\{f_w(i) = w_i\}] \\ \text{\hspace{2cm}} \Pi_{i \in \mathcal{L}_W} \delta_i^w((u_w)_i|(u_w)^{1:i-1}) \\ \text{\hspace{2.5cm}}   \Pi_{i \in R^w}P_{(U_w)_i|(U_w)^{1:{i-1}}} ((u_w)_i|(u_w)^{1:i-1})\big)$.\\
This implies that \\
$\mathbb{E}_{\mathbbm{C}}[{\mathbb{P}(U_w^{1:N}= u_w^{1:N}|  \mathbbm{C})}]\\ 
\text{\hspace{0.2cm}}= 2^{-|\mathcal{H}_W|} \Pi_{i \in \mathcal{L}_W} \delta_i^w((u_w)_i|(u_w)^{1:i-1}) \\ \text{\hspace{2.5cm}}  \Pi_{i \in R^w}P_{(U_w)_i|(U_w)^{1:{i-1}}} ((u_w)_i|(u_w)^{1:i-1})\big)$.\\
Similarly,  we give the private and public message bits in a portion of $U^{\mathcal{H}_{V|W}}$ and we give randomly chosen frozen bits with i.i.d.  uniform distribution in the remaining portion of it. Let $I^v$ be index set where we put private/public message bits in  $\mathcal{H}_{V|W}$ of the block we considered. Let the randomly chosen frozen bit function be  $f_v:\mathcal{H}_{V|W} - I_v \rightarrow \{0, 1\}$.  By encoding rule, we get \\
${\mathbb{P}(U_v^{1:N}= u_v^{1:N}|  \mathbbm{C}, W^{1:N} = w^{1:N}}) \\
\text{\hspace{0.2cm}}= 2^{-|I_v|} \Pi_{i \in \mathcal{H}_{V|W} - I} \mathbb{1}\{f_v(i) = v_i\}\\ \text{\hspace{1.2cm}} \Pi_{i \in \mathcal{L}_{V|W}} \delta_i^v((u_v)_i|w^{1:N} (u_v)^{1:i-1} ) \\ \text{\hspace{1.5cm}}   \Pi_{i \in R^v}P_{(U_v)_i|W^{1:N}(U_v)^{1:{i-1}}} ((u_v)_i|w^{1:N}(u_v)^{1:i-1})\big)$.\\
By taking expectation on both sides, by the independence of frozen bits and by the linearity of expectation, we get the following:\\
$\mathbb{E}_{\mathbbm{C}}[{\mathbb{P}(U_v^{1:N}= u_v^{1:N}|  \mathbbm{C}, W^{1:N} = w^{1:N}})]\\ 
\text{\hspace{0.2cm}}= 2^{-|I_v|} \Pi_{i \in \mathcal{H}_{V|W} - I_v} \mathbb{E}_{\mathbbm{C}}[\mathbb{1}\{f_v(i) = v_i\}]\\ \text{\hspace{1.2cm}} \Pi_{i \in \mathcal{L}_{V|W}} \delta_i^v((u_v)_i|w^{1:N} (u_v)^{1:i-1} ) \\ \text{\hspace{1.5cm}}   \Pi_{i \in R^v}P_{(U_v)_i|W^{1:N}(U_v)^{1:{i-1}}} ((u_v)_i|w^{1:N}(u_v)^{1:i-1})\big)$.\\
This implies that \\
$\mathbb{E}_{\mathbbm{C}}[{\mathbb{P}(U_v^{1:N}= u_v^{1:N}|  \mathbbm{C}, W^{1:N} = w^{1:N}})]\\\text{\hspace{0.2cm}}= 2^{-|\mathcal{H}_{V|W}|}  \Pi_{i \in \mathcal{L}_{V|W}} \delta_i^v((u_v)_i|(u_v)^{1:i-1} w^{1:N}) \\ \text{\hspace{1.5cm}}\Pi_{i \in R^v} P_{(U_v)_i| W^{1:N}(U_v)^{1:{i-1}} } ((u_v)_i|w^{1:N} (u_v)^{1:i-1} )$.\\
Similarly,  we give the private and public message bits in a portion of $U^{\mathcal{H}_{X|V}}$ and we give randomly chosen frozen bits with i.i.d.  uniform distribution in the remaining portion. Let $I^x$ be index set where we put private/public message bits in  $\mathcal{H}_{X|V}$ of the block we considered. Let the randomly chosen frozen bit function be  $f_x:\mathcal{H}_{X|V} - I_x \rightarrow \{0, 1\}$.  By encoding rule, we get \\
${\mathbb{P}(U_x^{1:N}= u_x^{1:N}|  \mathbbm{C}, V^{1:N} = v^{1:N}}) \\
\text{\hspace{0.2cm}}= 2^{-|I_x|} \Pi_{i \in \mathcal{H}_{X|V} - I_x} \mathbb{1}\{f_x(i) = x_i\}\\ \text{\hspace{1.2cm}} \Pi_{i \in \mathcal{L}_{X|V}} \delta_i^v((u_x)_i|v^{1:N} (u_x)^{1:i-1} ) \\ \text{\hspace{1.5cm}}   \Pi_{i \in R^x}P_{(U_x)_i|V^{1:N} (U_x)^{1:{i-1}}} ((u_x)_i|v^{1:N}(u_x)^{1:i-1})\big)$.\\
By taking expectation on both sides, by the independence of frozen bits and by the linearity of expectation, we get the following:\\
 $\mathbb{E}_{\mathbbm{C}}[{\mathbb{P}(U_x^{1:N}= u_x^{1:N}|  \mathbbm{C}, V^{1:N} = v^{1:N}})]\\ 
\text{\hspace{0.2cm}}= 2^{-|I_x|} \Pi_{i \in \mathcal{H}_{X|V} - I} \mathbb{E}_{\mathbbm{C}}[\mathbb{1}\{f_x(i) = x_i\}]\\ \text{\hspace{1.2cm}} \Pi_{i \in \mathcal{L}_{X|V}} \delta_i^x((u_x)_i|v^{1:N} (u_x)^{1:i-1} ) \\ \text{\hspace{1.5cm}}   \Pi_{i \in R^x}P_{(U_x)_i|V^{1:N} (U_x)^{1:{i-1}}} ((u_x)_i|v^{1:N}(u_x)^{1:i-1})\big)$.\\
This implies that \\
$\mathbb{E}_{\mathbbm{C}}[{\mathbb{P}(U_x^{1:N}= u_x^{1:N}|  \mathbbm{C}, V^{1:N} = v^{1:N}})]\\\text{\hspace{0.2cm}}= 2^{-|\mathcal{H}_{X|V}|}  \Pi_{i \in \mathcal{L}_{X|V}} \delta_i^x((u_x)_i|v^{1:N} (u_x)^{1:i-1} ) \\ \text{\hspace{1.5cm}}\Pi_{i \in R^x} P_{(U_x)_i|V^{1:N} (U_x)^{1:{i-1}} } ((u_x)_i|v^{1:N} (u_x)^{1:i-1} )$.\\
 By the chain-rule of conditional probability, we get \\
${\mathbb{P}(U_w^{1:N}= u_w^{1:N}, U_v^{1:N}= u_v^{1:N}, U_x^{1:N}= u_x^{1:N} |  \mathbbm{C})} . \\\text{\hspace{0.2cm}}= {\mathbb{P}(U_w^{1:N}= u_w^{1:N}|  \mathbbm{C})} \cdot {\mathbb{P}(U_v^{1:N}= u_v^{1:N}| \mathbbm{C} , W^{1:N} = w^{1:N} )}\cdot\\ \text{\hspace{2cm}} {\mathbb{P}(U_x^{1:N}= u_x^{1:N}| \mathbbm{C} , V^{1:N} = v^{1:N} )}.$\\
By taking expectations on the both the sides and by using the fact that the frozen bit functions $f_w$, $f_v$ and $f_x$ are independent, we get the following:\\
$\mathbb{E}_{\mathbbm{C}}[{\mathbb{P}(U_w^{1:N}= u_w^{1:N}, U_v^{1:N}= u_v^{1:N}, U_x^{1:N}= u_x^{1:N} |  \mathbbm{C})}]  \\ \text{\hspace{1.2cm}} = \mathbb{E}_{\mathbbm{C}}[{\mathbb{P}(U_w^{1:N}= u_w^{1:N}|  \mathbbm{C})}] \cdot \\ \text{\hspace{2cm}} \mathbb{E}_{\mathbbm{C}}[ {\mathbb{P}(U_v^{1:N}= u_v^{1:N}| \mathbbm{C} , W^{1:N} = w^{1:N} )}]\cdot\\ \text{\hspace{2.3cm}} \mathbb{E}_{\mathbbm{C}}[{\mathbb{P}(U_x^{1:N}= u_x^{1:N}| \mathbbm{C} , V^{1:N} = v^{1:N} )}].$\\
After substituting each of the three product terms on the right hand side, we finish the proof of part 1.\\
2.\\ Let $\mathcal{E}$ be the error event. Notice that the error occurs if and only if  there is an  error while decoding bit-channels $\mathcal{L}_{W} \cup I_j^w$ in $(U_w)^{1:N}$ for $j=1,2,3$ or $\mathcal{L}_{V|W} \cup I_j^v$ in $(U_v)^{1:N}$ for $j=1,3$ or $\mathcal{L}_{X|V} \cup I_1^x$ in $(U_x)^{1:N}$ in any of the blocks involved in the chaining construction.
Let us index the  blocks in chaining construction as $b=1,2, \ldots, k$.\\
 The error event of bit-channel $i$ of block $b$  for receivers $j=1,2$ or $3$ in the first layer will be as follows: 
\begin{align*}
 \mathcal{E}_{ij}^{wb} &=  \{(w^{1:N}, v^{1:N},x^{1:N}, y_j^{1:N})  \text{s of all the blocks $\Tilde{b} \in [k]$ }: \\&\text{\hspace{1cm}} P_{(U_w)_i|(U_w)^{1:{i-1}}Y_j^{1:N}}((u_w)_i +1 |{(u_w)}^{1:{i-1}} y_j^{1:N}) \\&\text{\hspace{1.2cm}}\geq P_{(U_w)_i|(U_w)^{1:{i-1}}Y_j^{1:N}}((u_w)_i|{(u_w)}^{1:{i-1}} y_j^{1:N}) \\ &\text{\hspace{2.5cm}} \text{holds for $(u_w^{1:N},y_j^{1:N})$ of block $b$}  \} .
 \end{align*}
 When there is only a single block, the error event of bit-channel $i$ for receivers $j=1,2$ or $3$ in the first layer will be as follows: 
\begin{align*}
 \mathcal{E}_{ij}^{w} &=  \{(w^{1:N}, v^{1:N},x^{1:N}, y_j^{1:N})  : \\&\text{\hspace{1cm}} P_{(U_w)_i|(U_w)^{1:{i-1}}Y_j^{1:N}}((u_w)_i +1 |{(u_w)}^{1:{i-1}} y_j^{1:N}) \\&\text{\hspace{1.2cm}}\geq P_{(U_w)_i|(U_w)^{1:{i-1}}Y_j^{1:N}}((u_w)_i|{(u_w)}^{1:{i-1}} y_j^{1:N})   \} .
 \end{align*}
  The error event of bit-channel $i$ of block $b$ for receivers $j=1$ or $3$ in the second layer  will be as follows: 
\begin{align*}
 \mathcal{E}_{ij}^{vb} &=  \{(w^{1:N}, v^{1:N},x^{1:N}, y_j^{1:N})  \text{s of all the blocks $\Tilde{b} \in [k]$ }: \\&\text{\hspace{0.2cm}} P_{(U_v)_i|W^{1:N} (U_v)^{1:{i-1}}Y_j^{1:N}}((u_v)_i +1 |w^{1:N} {(u_w)}^{1:{i-1}} y_j^{1:N}) \\&\geq\text{\hspace{0.2cm}} P_{(U_v)_i|W^{1:N} (U_v)^{1:{i-1}}Y_j^{1:N}}((u_v)_i|{w^{1:N} (u_w)}^{1:{i-1}} y_j^{1:N})\\ &\text{\hspace{2.2cm}} \text{holds for $(w^{1:N}, u_v^{1:N},y_j^{1:N})$ of block $b$} \} .
 \end{align*}
 When there is only a single block,  the error event of bit-channel $i$ for receivers $j=1$ or $3$ in the second layer  will be as follows: 
\begin{align*}
 \mathcal{E}_{ij}^{v} &=  \{(w^{1:N}, v^{1:N},x^{1:N}, y_j^{1:N}) : \\&\text{\hspace{0.2cm}} P_{(U_v)_i|W^{1:N} (U_v)^{1:{i-1}}Y_j^{1:N}}((u_v)_i +1|w^{1:N} {(u_w)}^{1:{i-1}} y_j^{1:N}) \\&\geq\text{\hspace{0.2cm}} P_{(U_v)_i|W^{1:N} (U_v)^{1:{i-1}}Y_j^{1:N}}((u_v)_i|{w^{1:N} (u_w)}^{1:{i-1}} y_j^{1:N})\} .
 \end{align*}
  The error event of bit-channel $i$ of block $b$ for receiver $j=1$ in the third layer will be as follows: 
\begin{align*}
 \mathcal{E}_{ij}^{xb} &=  \{(w^{1:N}, v^{1:N},x^{1:N}, y_j^{1:N}) \text{s of all the blocks $\Tilde{b} \in [k]$ }: \\&\text{\hspace{0.2cm}} P_{(U_x)_i|V^{1:N} (U_x)^{1:{i-1}}Y_j^{1:N}}((u_x)_i +1 |v^{1:N} {(u_x)}^{1:{i-1}} y_j^{1:N}) \\&\geq\text{\hspace{0.2cm}} P_{(U_x)_i|V^{1:N} (U_x)^{1:{i-1}}Y_j^{1:N}}((u_x)_i|{v^{1:N} (u_x)}^{1:{i-1}} y_j^{1:N})\\ &\text{\hspace{2.2cm}} \text{holds for $(v^{1:N}, u_x^{1:N},y_j^{1:N})$ of block $b$} \} .
 \end{align*}
When there is only a single block, the error event of bit-channel $i$ for receiver $j=1$ in the third layer will be as follows: 
\begin{align*}
 \mathcal{E}_{ij}^{x} &=  \{(w^{1:N}, v^{1:N},x^{1:N}, y_j^{1:N}) : \\&\text{\hspace{0.2cm}} P_{(U_x)_i|V^{1:N} (U_x)^{1:{i-1}}Y_j^{1:N}}((u_x)_i + 1|v^{1:N} {(u_x)}^{1:{i-1}} y_j^{1:N}) \\&\geq\text{\hspace{0.2cm}} P_{(U_x)_i|V^{1:N} (U_x)^{1:{i-1}}Y_j^{1:N}}((u_x)_i|{v^{1:N} (u_x)}^{1:{i-1}} y_j^{1:N})\} .
 \end{align*}
  We define $\mathcal{E}_j^{wb} = \cup_{i\in I_j^w \cup \mathcal{L}_W}\mathcal{E}_{ij}^{wb}$  for $j=1,2,3$,  $\mathcal{E}_j^{vb} = \cup_{i\in I_j^v \cup \mathcal{L}_{V|W}}\mathcal{E}_{ij}^{vb}$ for $j=1,3$ and  $\mathcal{E}_j^{xb} = \cup_{i\in I_j^x \cup \mathcal{L}_{X|V}}\mathcal{E}_{ij}^{xb}$ for $j=1$.\\
  We define $\mathcal{E}_j^{w} = \cup_{i\in I_j^w \cup \mathcal{L}_W}\mathcal{E}_{ij}^{w}$  for $j=1,2,3$,  $\mathcal{E}_j^{v} = \cup_{i\in I_j^v \cup \mathcal{L}_{V|W}}\mathcal{E}_{ij}^{v}$ for $j=1,3$ and  $\mathcal{E}_j^{x} = \cup_{i\in I_j^x \cup \mathcal{L}_{X|V}}\mathcal{E}_{ij}^{x}$ for $j=1$.\\
 We define $\mathcal{E}_1^b = \mathcal{E}_1^{wb} \cup \mathcal{E}_1^{vb} \cup \mathcal{E}_1^{xb}  $,  $\mathcal{E}_2^b = \mathcal{E}_2^{wb}  $ and  $\mathcal{E}_3^b = \mathcal{E}_3^{wb} \cup \mathcal{E}_3^{vb}$ for each block $b$. \\
  We define $\mathcal{E}_{1s} = \mathcal{E}_1^{w} \cup \mathcal{E}_1^{v} \cup \mathcal{E}_1^{x}  $,  $\mathcal{E}_{2s} = \mathcal{E}_2^{w}  $ and  $\mathcal{E}_{3s} = \mathcal{E}_3^{w} \cup \mathcal{E}_3^{v}$ for each block $b$. \\
 We define $\mathcal{E}_j =      \cup_{b=1}^{k} \mathcal{E}_j^b$, which will be error event for receiver-$j$,  where $j=1,2,3$.\\
 Therefore the overall error event $\mathcal{E} = \cup_{j=1}^3 \mathcal{E}_j$.  By union bound, we the following identity:\\
$ \mathbb{P}(\mathcal{E}|\mathbbm{C}) \leq \sum_{j=1}^{3} \sum_{b=1}^{k} \mathbb{P}(\mathcal{E}_j^b|\mathbbm{C}) $ .\\
By taking expectation on both the sides and also by applying linearity of expectation, we get \\
\begin{equation}\label{eq:5}
 \mathbb{E}_{\mathbbm{C}}[\mathbb{P}(\mathcal{E}|\mathbbm{C})] \leq \sum_{j=1}^{3} \sum_{b=1}^{k} \mathbb{E}_{\mathbbm{C}}[\mathbb{P}(\mathcal{E}_j^b|\mathbbm{C})].
 \end{equation}
Let $Q_{((U_w)^{1:N} (U_v)^{1:N} (U_x)^{1:N})}$ be the measure on $((U_w)^{1:N}(U_v)^{1:N} (U_x)^{1:N})$  as follows:\\
 $Q_{((U_w)^{1:N} (U_v)^{1:N}(U_x)^{1:N}}( u_w^{1:N}, u_v^{1:N}, u_x^{1:N})  \\\text{\hspace{0.2cm}}= Q_{(U_w)^{1:N}}(u_w^{1:N}) Q_{(U_v)^{1:N}| W^{1:N}= w^{1:N}}(u_w^{1:N})   \\ \text{\hspace{2.7cm}}  Q_{(U_x)^{1:N}| V^{1:N}= v^{1:N}}(u_v^{1:N}) \\\text{\hspace{0.2cm}}=
\big(2^{-|\mathcal{H}_W|} \Pi_{i \in \mathcal{L}_W} \delta_i^w((u_w)_i|(u_w)^{1:i-1}) \\ \text{\hspace{2.7cm}}   \Pi_{i \in R^w}P_{(U_w)_i|(U_w)^{1:{i-1}}} ((u_w)_i|(u_w)^{1:i-1})\big)\cdot
\text{\hspace{0.7cm}}
\big(2^{-|\mathcal{H}_{V|W}|}  \Pi_{i \in \mathcal{L}_{V|W}} \delta_i^v((u_v)_i|w^{1:N} (u_v)^{1:i-1} ) \\ \text{\hspace{1.5cm}}\Pi_{i \in R^v} P_{(U_v)_i|W^{1:N} (U_v)^{1:{i-1}} }  ( (u_v)_i|w^{1:N} (u_v)^{1:i-1} )\big)\cdot
\\\text{\hspace{0.7cm}}
\big(2^{-|\mathcal{H}_{X|V}|} \Pi_{i \in \mathcal{L}_{X|V}} \delta_i^x((u_x)_i|v^{1:N} (u_x)^{1:i-1} ) \\ \text{\hspace{1.5cm}}\Pi_{i \in R^x} P_{(U_x)_i|V^{1:N} (U_x)^{1:{i-1} }} ((u_x)_i|v^{1:N} (u_x)^{1:i-1}  ) \big).$\\
Note that $P_{(U_w)^{1:N} (U_v)^{1:N} (U_x)^{1:N}}$ is the measure induced when $(W^{1:N}, V^{1:N}, X^{1:N})$ is  i.i.d. according to the distribution $p(w)p(v|w)p(x|v)$.\\
From Lemma \ref{lemma:3}, we have\\
$ ||P_{(U_w)^{1:N}(U_v)^{1:N} (U_x)^{1:N}} - \\ \text{\hspace{2.5cm}} Q_{(U_w)^{1:N}(U_v)^{1:N} (U_x)^{1:N}}
|| = O(2^{-N^{\beta'}}).$\\
where $\beta' < \beta $.\\
\remove{
By using the results from \cite{chou2} and  \cite{honda}, note that we get the following:
\begin{equation}\label{eq:4}
|| P_{(U_w)^{1:N}} - Q_{(U_w)^{1:N}} || = O(2^{-N^{\beta'}}),
\end{equation}
\begin{equation}\label{eq:6}
|| P_{(U_v)^{1:N}|W^{1:N} = w^{1:N}} - Q_{(U_v)^{1:N}|W^{1:N} = w^{1:N}} || = O(2^{-N^{\beta'}}),
\end{equation}
for each $w^{1:N} \in \{0,1\}^{N}$,
\begin{equation}\label{eq:7}
|| P_{(U_x)^{1:N}|V^{1:N} = v^{1:N}} - Q_{(U_x)^{1:N}|V^{1:N} = v^{1:N}} || = O(2^{-N^{\beta'}}),
\end{equation}
for each $v^{1:N} \in \{0,1\}^{N}$, 
for some $\beta' < \beta < 0.5$.
Now the total variation distance
\begin{align*}
&||P_{(U_w)^{1:N} (U_v)^{1:N} (U_x)^{1:N}} - Q_{(U_w)^{1:N} (U_v)^{1:N} (U_x)^{1:N}}|| \\
&\stackrel{(a)}\leq ||P_{(U_w)^{1:N} (U_v)^{1:N} } - Q_{(U_w)^{1:N} (U_v)^{1:N} }|| + O(2^{-N^{\beta'}})\\
&\stackrel{(b)}\leq ||P_{(U_w)^{1:N}}   - Q_{(U_w)^{1:N}}  || + O(2^{-N^{\beta'}}) +  O(2^{-N^{\beta'}})\\
&\leq  O(2^{-N^{\beta'}}) +  O(2^{-N^{\beta'}})  +  O(2^{-N^{\beta'}})\\
&= O(2^{-N^{\beta'}}).
\end{align*}

Identity (a) follows from the application of Lemma \ref{lemma:2}  where $((U_w)^{1:N}, (U_v)^{1:N})$ takes role of $X$ and $(U_x)^{1:N}$ takes the role of $Y$. We also use the fact that $W^{1:N}$, $V^{1:N}$ and $X^{1:N}$ form a markov chain and use equation (\ref{eq:7}). \\
Identity (b) follows from the application of Lemma \ref{lemma:2} where $(U_w)^{1:N}$ takes role of $X$ and $(U_v)^{1:N}$ takes the role of $Y$. We also use equation (\ref{eq:6}).\\}
$\mathbb{P}(\mathcal{E}_j^b|\mathbbm{C})\\ \text{\hspace{0.4cm}} = \sum_{ ((u_w)^{1:N} , (u_v)^{1:N}, (u_x)^{1:N},y_j^{1:N}) \text{s of all blocks $[k]$)} \in \mathcal{E}_j^b} \\ \text{\hspace{1.5cm}} \mathbb{P} ( \cap_{\tilde{b} \in [k]} (U_w^{1:N}= u_w^{1:N}, U_v^{1:N}= u_v^{1:N}, \\\text{\hspace{2cm}} U_x^{1:N}= u_x^{1:N}, Y_j^{1:N} = y_j^{1:N}  \text{ of block $\tilde{b}$} )| \mathbbm{C}).$\\ 
From the definitions of $\mathcal{E}_j^b$ and $\mathcal{E}_{js}$, we get\\
$\mathbb{P}(\mathcal{E}_j^b|\mathbbm{C})\\ \text{\hspace{0.4cm}} =  \sum_{(((u_w)^{1:N} , (u_v)^{1:N}, (u_x)^{1:N},y_j^{1:N}) \text{ of block $b$}) \in \mathcal{E}_{js}} \\ \text{\hspace{1cm}}\sum_{ (((u_w)^{1:N} , (u_v)^{1:N}, (u_x)^{1:N},y_j^{1:N}) \text{s of  blocks $[k]- \{b\}$)} } \\ \text{\hspace{1.5cm}} \mathbb{P} ( \cap_{\tilde{b} \in [k]} (U_w^{1:N}= u_w^{1:N}, U_v^{1:N}= u_v^{1:N}, \\\text{\hspace{2cm}} U_x^{1:N}= u_x^{1:N}, Y_j^{1:N} = y_j^{1:N}  \text{ of block $\tilde{b}$} )| \mathbbm{C}).$\\
 By marginalizing over \\
 $(U_w^{1:N}, U_v^{1:N}, U_x^{1:N}, Y_j^{1:N})\text{s of  blocks $[k] - \{b\}$} $, we now get 
 
 $\mathbb{P}(\mathcal{E}_j^b|\mathbbm{C})\\ \text{\hspace{0.4cm}} =  \sum_{(((u_w)^{1:N} , (u_v)^{1:N}, (u_x)^{1:N},y_j^{1:N}) \text{ of block $b$}) \in \mathcal{E}_{js}}  \\ \text{\hspace{1.5cm}} \mathbb{P} ((U_w^{1:N}= u_w^{1:N}, U_v^{1:N}= u_v^{1:N}, \\\text{\hspace{2cm}} U_x^{1:N}= u_x^{1:N}, Y_j^{1:N} = y_j^{1:N}  \text{ of block ${b}$} )| \mathbbm{C}).$\\
 By chain rule of condition probability and also by the fact that  \\
 $\mathbb{P}(Y_j^{1:N} = y_j^{1:N} \text{ of block $b$} |X^{1:N} = x^{1:N} \text{ of block $b$, } \mathbbm{C}) \\ \text{\hspace{2cm}}= \Pi_{i=1}^N p(y_{ji} |x_i), $\\
 we will have the following:\\
$\mathbb{P}(\mathcal{E}_j^b|\mathbbm{C})\\ \text{\hspace{0.4cm}} =  \sum_{(((u_w)^{1:N} , (u_v)^{1:N}, (u_x)^{1:N},y_j^{1:N}) \text{ of block $b$}) \in \mathcal{E}_{js}}  \\ \text{\hspace{1.5cm}}  \mathbb{P}((U_w^{1:N}= u_w^{1:N}, U_v^{1:N}= u_v^{1:N}, \\\text{\hspace{2cm}} U_x^{1:N}= u_x^{1:N} \text{ of block $b$}|\mathbbm{C}) \Pi_{i=1}^N p(y_{ji} |x_i).$ \\
In the term $\Pi_{i=1}^N p_l(y_{ji} |x_i)$ here,  notice that $x^{1:N}$ vector is corresponding to block $b$, which means it is obtained by applying polar transform to $(u_x)^{1:N}$ vector corresponding to block $b$ and also  $y_j^{1:N}$ vector is corresponding to block $b$.\\
By taking expectation on both the sides and by the linearity of expectation, we get the following:\\
$\mathbb{E}_{\mathbbm{C}}[\mathbb{P}(\mathcal{E}_j^b|\mathbbm{C})]\\ \text{\hspace{0.3cm}}= \sum_{ ((u_w)^{1:N} , (u_v)^{1:N}, (u_x)^{1:N} ,y_j^{1:N} ) \text{ of block ${b}$) } \in \mathcal{E}_{js}} \\ \text{\hspace{1.0cm}} \mathbb{E}_{\mathbbm{C}}[ \mathbb{P} (U_w^{1:N}= u_w^{1:N}, U_v^{1:N}= u_v^{1:N},\\\text{\hspace{4cm}} U_x^{1:N}= u_x^{1:N} \text{ of block $b$} | \mathbbm{C})]\\\text{\hspace{4.5cm}}\Pi_{i=1}^Np(y_{ji}|x_i)\\
\text{\hspace{0.3cm}}\stackrel{(a)}=\sum_{ ((u_w)^{1:N} , (u_v)^{1:N}, (u_x)^{1:N} ,y_j^{1:N})\text{ of block ${b}$) } \in \mathcal{E}_{js}} \\ \text{\hspace{0.8cm}}
Q_{(U_w)^{1:N} (U_v)^{1:N}(U_x)^{1:N}}( u_w^{1:N} u_v^{1:N} u_x^{1:N}) \Pi_{i=1}^Np(y_{ji}|x_i)\\
\text{\hspace{0.3cm}}= Q_{(U_w)^{1:N} (U_v)^{1:N}(U_x)^{1:N} Y_j^{1:N}}(\mathcal{E}_{js})\\
\text{\hspace{0.2cm}}\leq ||  P_{(U_w)^{1:N} (U_v)^{1:N}(U_x)^{1:N}Y_j^{1:N}} - \\
\text{\hspace{3.3cm}} Q_{(U_w)^{1:N} (U_v)^{1:N}(U_x)^{1:N}Y_j^{1:N}}|| \\\text{\hspace{3.6cm}}+ P_{(U_w)^{1:N} (U_v)^{1:N}(U_x)^{1:N}Y_j^{1:N}}(\mathcal{E}_{js})\\
\text{\hspace{0.2cm}} \stackrel{(b)}= ||  P_{(U_w)^{1:N} (U_v)^{1:N}(U_x)^{1:N}} -  Q_{(U_w)^{1:N} (U_v)^{1:N}(U_x)^{1:N}}|| \\\text{\hspace{3.6cm}}+ P_{(U_w)^{1:N} (U_v)^{1:N}(U_x)^{1:N}Y_j^{1:N}}(\mathcal{E}_{js})\\
\text{\hspace{0.3cm}}=O(2^{-N^{\beta'}}) + P_{(U_w)^{1:N} (U_v)^{1:N}(U_x)^{1:N}Y_j^{1:N}}(\mathcal{E}_{js})$.\\
Identity (a) follows from part 1. Identity (b) follows from Lemma \ref{lemma:4}.\\
For receiver-$1$, that is $j=1$, we get   \\ $\mathbb{E}_C[\mathbb{P}(\mathcal{E}_1^b)|\mathbbm{C}]\\
\text{\hspace{0.2cm}}= O(2^{-N^{\beta'}}) + P_{(U_w)^{1:N} (U_v)^{1:N}(U_x)^{1:N}Y_1^{1:N}}(\mathcal{E}_{1s})\\\text{\hspace{0.2cm}} \stackrel{(a)}\leq O(2^{-N^{\beta'}}) + P_{(U_w)^{1:N} (U_v)^{1:N}(U_x)^{1:N}}(\mathcal{E}_1^{w}) \\ \text{\hspace{1.6cm}}+ P_{(U_w)^{1:N} (U_v)^{1:N}(U_x)^{1:N}Y_1^{1:N}}(\mathcal{E}_1^{v}) \\ \text{\hspace{1.6cm}}+ P_{(U_w)^{1:N} (U_v)^{1:N}(U_x)^{1:N}Y_1^{1:N}}(\mathcal{E}_1^{x}) \\
\text{\hspace{0.2cm}} \stackrel{(b)}\leq O(2^{-N^{\beta'}}) \\ \text{\hspace{1.6cm}}+ \sum_{i \in \mathcal{L}_W \cup I^w_1} P_{(U_w)^{1:N} (U_v)^{1:N}(U_x)^{1:N}Y_1^{1:N}}(\mathcal{E}_{i1}^{w}) \\ \text{\hspace{1.6cm}}+ \sum_{i \in \mathcal{L}_{V|W} \cup I^v_1} P_{(U_w)^{1:N} (U_v)^{1:N}(U_x)^{1:N}Y_1^{1:N}}(\mathcal{E}_{i1}^{v}) \\ \text{\hspace{1.6cm}}+\sum_{i \in \mathcal{L}_{X|V} \cup I^x_1} P_{(U_w)^{1:N} (U_v)^{1:N}(U_x)^{1:N}Y_1^{1:N}}(\mathcal{E}_{i1}^{x}) \\
\text{\hspace{0.2cm}}  \leq O(2^{-N^{\beta'}}) + \sum_{i \in \mathcal{L}_W \cup I^w_1} Z((U_w)_i | (U_w)^{1:i-1}Y_1^{1:N}) \\ \text{\hspace{1.6cm}}+ \sum_{i \in \mathcal{L}_{V|W} \cup I^v_1} Z((U_v)_i |W^{1:N} (U_v)^{1:i-1}Y_1^{1:N}) \\ \text{\hspace{1.6cm}}+\sum_{i \in \mathcal{L}_{X|V} \cup I^x_1} Z((U_x)_i |V^{1:N} (U_x)^{1:i-1}Y_1^{1:N})\\
 \text{\hspace{0.2cm}} \leq  O(2^{-N^{\beta'}}) + N2^{-N^{\beta}} + N2^{-N^{\beta}} + N2^{-N^{\beta}} \\
\text{\hspace{0.2cm}}  \leq O( 2^{-N^{\beta'}}).$ \\
 Identity (a) follows from the definition of $\mathcal{E}_{1s}$ and union bound. Identity (b) follows from the definition of $\mathcal{E}_1^{w}$, $\mathcal{E}_1^{v}$, $\mathcal{E}_1^{x}$ and union bound.\\
 For receiver-$2$, that is $j=2$, we get  \\ $\mathbb{E}_C[\mathbb{P}(\mathcal{E}_2^b)|\mathbbm{C}]\\
\text{\hspace{0.2cm}} = O(2^{-N^{\beta'}}) + P_{(U_w)^{1:N} (U_v)^{1:N}(U_x)^{1:N}Y_2^{1:N}}(\mathcal{E}_{2s})\\\text{\hspace{0.2cm}}  \stackrel{(a)}\leq O(2^{-N^{\beta'}}) + P_{(U_w)^{1:N} (U_v)^{1:N}(U_x)^{1:N}Y_2^{1:N}}(\mathcal{E}_2^{w})  \\
\text{\hspace{0.2cm}} \stackrel{(b)} \leq O(2^{-N^{\beta'}}) + \sum_{i \in \mathcal{L}_W \cup I^w_2} P_{(U_w)^{1:N} (U_v)^{1:N}(U_x)^{1:N}Y_2^{1:N}}(\mathcal{E}_{i2}^{w}) \\
\text{\hspace{0.2cm}}  \leq O(2^{-N^{\beta'}}) + \sum_{i \in \mathcal{L}_W \cup I^w_2} Z((U_w)_i | (U_w)^{1:i-1}Y_2^{1:N}) \\
 \text{\hspace{0.2cm}} \leq O( 2^{-N^{\beta'}}) + N2^{-N^{\beta}}\\
\text{\hspace{0.2cm}}   = O( 2^{-N^{\beta'}}).$\\
  Identity (a) follows from the definition of $\mathcal{E}_{2s}$. Identity (b) follows from the definition of $\mathcal{E}_2^{w}$ and union bound.\\
 For receiver-$3$, that is $j=3$, we get  \\ $\mathbb{E}_C[\mathbb{P}(\mathcal{E}_3^b)|\mathbbm{C}]\\
\text{\hspace{0.2cm}} = O(2^{-N^{\beta'}}) + P_{(U_w)^{1:N} (U_v)^{1:N}(U_x)^{1:N}Y_3^{1:N}}(\mathcal{E}_{3s})\\ \text{\hspace{0.2cm}} \stackrel{(a)}\leq O(2^{-N^{\beta'}}) + P_{(U_w)^{1:N} (U_v)^{1:N}(U_x)^{1:N}Y_3^{1:N}}(\mathcal{E}_3^{w}) \\ \text{\hspace{1.6cm}}+ P_{(U_w)^{1:N} (U_v)^{1:N}(U_x)^{1:N}Y_3^{1:N}}(\mathcal{E}_3^{v}) \\
\text{\hspace{0.2cm}} \stackrel{(b)} \leq O(2^{-N^{\beta'}}) + \sum_{i \in \mathcal{L}_W \cup I^w_3} P_{(U_w)^{1:N} (U_v)^{1:N}(U_x)^{1:N}Y_3^{1:N}}(\mathcal{E}_{i3}^{w}) \\ \text{\hspace{1.6cm}}+ \sum_{i \in \mathcal{L}_{V|W} \cup I^v_3} P_{(U_w)^{1:N} (U_v)^{1:N}(U_x)^{1:N}Y_3^{1:N}}(\mathcal{E}_{i3}^{v})  \\
\text{\hspace{0.2cm}}  \leq O(2^{-N^{\beta'}}) + \sum_{i \in \mathcal{L}_W \cup I^w_3} Z((U_w)_i | (U_w)^{1:i-1}Y_3^{1:N}) \\ \text{\hspace{1.6cm}}+ \sum_{i \in \mathcal{L}_{V|W} \cup I^v_3} Z((U_v)_i |W^{1:N} (U_v)^{1:i-1}Y_3^{1:N}) \\
\text{\hspace{0.2cm}}  \leq  O(2^{-N^{\beta'}}) + N2^{-N^{\beta}} + N2^{-N^{\beta}}  \\
\text{\hspace{0.2cm}}  \leq O( 2^{-N^{\beta'}}).$ \\
  Identity (a) follows from the definition of $\mathcal{E}_{3s}$ and union bound. Identity (b) follows from the definition of $\mathcal{E}_3^{w}$, $\mathcal{E}_3^{v}$ and union bound.\\
 From equation (\ref{eq:5}), the overall average probability of error will become $O( k 2^{-N^{\beta'}})$. This concludes the proof of part 2. Hence the proof of Theorem \ref{theorem:1}. \qed\\
Both encoding and decoding complexities will become $O(N \log N )$ per block \cite{honda}.

 We have given the code-construction for the case where $|\mathcal{X}|=|\mathcal{V}|=|\mathcal{W}|=2$. If any of these alphabets have arbitrary sizes, we can adapt multi-level polar code construction technique. Let $|\mathcal{X}|$= $\Pi_{j=1}^m p_j$ $|\mathcal{V}|$= $\Pi_{j=1}^l q_j$ $|\mathcal{W}|$= $\Pi_{j=1}^k r_j$ where $\{r_j\}$, $\{q_j\}$ and $\{p_j\}$ are prime factors of $\mathcal{W}$, $\mathcal{V}$ and $\mathcal{X}$, respectively. Then random variables $W, V$ and $X$ can be represented by random vectors $(W_1,\ldots,W_k)$, $(V_1,\ldots,V_l)$ and $(X_1,\ldots,X_m)$ where $W_j$, $V_j$ and $X_j$ are supported over the set $\{0,1,\ldots ,r_j-1\}$, $\{0,1,\ldots q_j-1\}$ and  $\{0,1,\ldots p_j-1\}$, respectively.  By chain-rule of entropy, we get $
 H(W,V,X) = \Sigma_{j=1}^k H(W_j|W^{1:j-1})+ \Sigma_{j=1}^l H(V_j|W V^{1:j-1})  + \Sigma_{j=1}^m H(X_j|W V X^{1:j-1})$. 
 We can use the polarization for prime alphabets for each term in the above identity and use a polar code construction technique with an appropriate successive cancellation decoder~\cite{sasoglu1}, \cite{sasoglu2}. The key ideas in the analysis of the probability of error we provided for the binary case still apply to the coding method for larger alphabets and can be easily extended.
 \subsection{Extension: receiver-$1$ requires only $M_1$}
  For a $(2^{NR_0}, 2^{NR_1}, N)$ code of a setting with degraded messages sets, the converse proof of the capacity region just uses the fact that $H(M_1|Y_1^{1:N})$, $H(M_0|Y_2^{1:N})$
and $H(M_0|Y_3^{1:N})$ are $o(N)$~\cite{nair }. We do not have to use the stronger fact that  $H(M_1, M_0|Y_1^{1:N})$ is $o(N)$ to complete the converse proof. This means that the same proof becomes the converse proof of the capacity region for the problem when  receiver-$1$ requires to recover only $M_1$. Hence, the capacity region does not enlarge and remains the same. So the same polar coding method can be used to achieve all rate pairs inside the capacity region.
\end{section}

\begin{section}{Conclusion}\label{conclusion}
We considered the problem of achieving the rates in the capacity region of a discrete memoryless multi-level 3-receiver broadcast channel with degraded message sets  through polar coding. The problem  is to transmit a public message to all the receivers and a private message intended for receiver-$1$. Our motivation for this problem is   due to a file transfer application in a client-server network that has three clients, where this setting can be applied.
We give a new two-level chaining construction to achieve all the points in the capacity region without time-sharing. We also gave a detailed analysis of the probability of decoding error for constructed coding scheme.  We showed that the capacity of the broadcast channel does not enlarge,  even when receiver-$1$ requires to recover only its private message. Hence, we can use the same polar coding strategy to achieve the capacity under this setting.
\end{section}

\begin{section}{Acknowledgement}\label{ack}
This work was supported in part by National Science Foundation Grants CCF-1415109 and CCF-1619053.
\end{section}

\end{document}